\documentclass{emulateapj}
\usepackage{graphicx, amsmath, amssymb, amsthm}

\makeatother

\newcommand\Ls{L_\odot}
\newcommand\Ms{M_\odot}

\newcommand\Zs{Z_\odot}

\begin{document}
\title{The Destruction of Cosmological Minihalos by Primordial Supernovae}
\author{Daniel Whalen\altaffilmark{1,4}, Bob van Veelen\altaffilmark{2},
Brian W. O'Shea\altaffilmark{1,3}, \& Michael L. Norman \altaffilmark{4}}
\altaffiltext{1}{Applied Physics (X-2), Los Alamos National
Laboratory}
\altaffiltext{2}{Astronomical Institute Utrecht, Princetonplein 5, Utrecht, The Netherlands}
\altaffiltext{3}{Theoretical Astrophysics (T-6), Los Alamos National
Laboratory}
\altaffiltext{4}{Center for Astrophysics and Space Sciences,
University of California at San Diego, La Jolla, CA 92093, U.S.A.
Email: dwhalen@lanl.gov}

\begin{abstract} 

We present numerical simulations of primordial supernovae in cosmological minihalos at 
$z \sim$ 20.  We consider Type II supernovae, hypernovae, and pair instability supernovae 
(PISN) in halos from 6.9 $\times$ 10$^5$ - 1.2 $\times$ 10$^7$ $\Ms$, those in which 
Population III stars are expected to form via H$_2$ cooling.  Our simulations are novel 
in that they are the first to follow the evolution of the blast from a free expansion on 
spatial scales of 10$^{-4}$ pc until its approach to pressure equilibrium in the relic 
\ion{H}{2} region of the progenitor, $\sim$ 1000 pc.  Our models include nine-species 
primordial chemistry together with all atomic H and He cooling processes, inverse Compton 
cooling and free-free emission.  The supernovae evolve along two evolutionary paths according 
to whether they explode in \ion{H}{2} regions or neutral halos. Those in \ion{H}{2} regions 
first expand adiabatically and then radiate strongly upon collision with baryons ejected from 
the halo during its photoevaporation by the progenitor.  Explosions in neutral halos promptly 
emit most of their kinetic energy as x-rays, but retain enough momentum to seriously disrupt 
the halo.  We find that the least energetic of the supernovae are capable of destroying halos 
$\lesssim$ 10$^7$ $\Ms$, while a single PISN can destroy even more massive halos.  Blasts in 
\ion{H}{2} regions disperse heavy elements into the IGM, but neutral halos confine the 
explosion and its metals.  Primordial supernova remnants develop dynamical instabilities at 
early times capable of enriching up to 10$^6$ $\Ms$ of baryons with metals to levels $\gtrsim$ 
0.01 $\Zs$, well above that required for low-mass star formation.  In \ion{H}{2} regions, a 
prompt second generation of stars may form in the remnant at radii of 100 - 200 pc in the halo. 
Explosions confined by large halos instead recollapse, with infall rates in excess of 10$^{-2}$ 
$\Ms$ yr$^{-1}$ that heavily contaminate their interior.  This fallback may either fuel massive 
black hole growth at very high redshifts or create the first globular cluster with a radius of 
10 - 20 pc at the center of the halo.  Our findings allow the possibility that the first primitive 
galaxies formed sooner, with greater numbers of stars and distinct chemical abundance patterns,
than in current models.
 
\end{abstract}


\section{Introduction}

Numerical studies indicate that primordial stars form in the first pregalactic objects to reach 
masses of $\sim$ 5 $\times$ 10$^{5}$ $M_{\odot}$ at $z \sim$ 20 - 50 \citep{bcl99,abn00,abn02,gao07}.
The models suggest that Population III stars form in isolation (one per halo) and that they are 
likely very massive, from 100 to 500 $M_{\odot}$.  More recent work surveying a larger sample of halos 
\citep{oshea07b} extends the lower mass limit of primordial stars down to 15 $M_{\odot}$.  Very massive 
stars easily photoionize their halos, typically in a few hundred kyr with final \ion{H}{2} region radii 
of 2500 - 5000 pc while \ion{H}{2} regions of less massive stars are confined to very small radii ($<$ 
10$^{-3}$ pc) by the high central densities deep in the halo \citep{wan04,ket04,abs06,awb07,jet07,yet07,wa07a,wa07b}.  
In the latter case, high recombination rates and pressure from overlying matter due to the depth of 
the halo dark matter potential well halts the expansion of the heated bubble \citep{ftb90}.  Unlike 
massive stars in the Galaxy today, little mass loss occurs in Population III stars because there are 
no line-driven winds from their pristine atmospheres \citep{bhw01,kud00,vkl01}, although mixing in 
rapidly rotating stars may cause some mass loss later in their lives \citep{met07}.  Consequently, 
winds do not clear gas from cosmological halos in the first generation of stars as they do from 
molecular cloud cores today.  However, if the ionization front (I-front) of the star breaks out of 
the halo it will sweep more than half of the baryons interior to the virial radius into a thick 
shell $\sim$ 100 pc in radius by the end of the life of the star \citep{wan04,ket04}.  The diffuse 
gas enclosed by the shell is uniform and at densities far lower than in the original halo, 0.1 - 1 
cm$^{-3}$ instead of more than 10$^{10}$ cm$^{-3}$.

The ultimate fate of a Population III star depends on its mass at the end of its main sequence 
lifetime (MSL).  Those lying between 10 $M_{\odot}$ and 40 $M_{\odot}$ die in Type II supernova 
explosions, ejecting roughly 20\% of their mass into the intergalactic medium (IGM) as metals and 
leaving behind a compact remnant.  Stars between 140 $M_{\odot}$ and 260 $M_{\odot}$ explode in 
extremely energetic pair instability supernovae (PISN) hundreds of times more powerful \citep{hw02}.  
As much as 50\% of the mass of the star is dispersed as heavy elements into the IGM in such events, 
with no compact remnant.  However, these predictions are based on one-dimensional non-rotating 
stellar evolution models that make a range of assumptions (e.~g. mixing parameters) and 
should be considered qualitative, rather than quantitative, in nature.  How far elements are 
expelled from the star depends strongly on the state of the halo prior to the explosion.  Supernova 
(SN) remnants in halos completely ionized by the star typically expand to half the diameter of the H 
II region.  The shock front comes into pressure equilibrium with the hot recombining gas more quickly 
than if the blast propagates in a neutral medium of comparable density \citep{get07}. 

Supernovae in neutral halos, which result when low-mass Population III stars fail to ionize them, 
evolve quite differently.  Because no winds evacuate gas from the center of the halo, the blast 
encounters circumstellar densities several orders of magnitude greater than in OB associations in 
the Galaxy, with much higher energy losses at very early times.  Since the blast energy is much 
greater than the binding energy of the gas in the halo, it is often supposed that PISN easily 
destroyed halos.  Here, destruction of the halo means complete expulsion of its gas, not dispersal 
of the dark matter itself, which remains mostly undisturbed.  However, as \citet{ky05} point out, 
this argument ignores radiative losses from the shock in the large densities deep in the halo.  
They find that bremsstrahlung and inverse Compton cooling (IC) radiate away the explosion's energy 
before it even sweeps up its own mass in ambient gas, quenching the blast and preventing any 
dissemination of metals into the halo.  Explosions with hundreds of times the binding energy
may be necessary to disperse more massive halos under these conditions.

However, a difficulty with this and other studies of primordial supernovae in cosmological halos 
is how the explosion itself is initialized.  This is typically done by depositing the energy 
of the blast as thermal energy on the mesh and allowing pressure gradients to launch shocked flows 
into their surroundings \citep{byh03,ky05,get07}.  This is partly justified for explosions in \ion{H}{2} 
regions whose central densities are quite low and in which early radiative losses from the remnant 
are minimal.  Too much energy loss from the early remnant would be ignored by this approach for SN 
in neutral halos, which cool strongly from very early times due to high ambient densities.

In fact, explosions initialized by thermal pulses in neutral halos cool before any of their energy 
can be converted into motion \citep{ky05}.  When the energy of the blast is deposited as heat to 
central mesh zones, their temperatures skyrocket to tens of billions of degrees K, driving extreme 
bremsstrahlung and IC losses before pressure gradients can accelerate any flows.  These initial 
conditions do not properly represent the free expansion that actually erupts through the stellar 
envelope and whose momentum cannot be radiated away.  The early stages of the explosion in a 
nearly undisturbed halo are much better modeled by a cold free expansion in which the energy 
of the blast is kinetic rather than thermal.  Furthermore, supernovae modeled in cosmological 
density fields with smoothed particle hydrodynamics methods \citep{byh03,get07} also lack the 
numerical resolution to capture the early hydrodynamic behavior of the blast wave.  Dynamical 
instabilities might manifest that lead to early mixing of heavy elements in the ejecta with 
ambient halo material on much earlier time scales than in current simulations.   

How easily neutral halos are destroyed by primordial supernovae and how metals mix with the early
IGM thus remains unclear.  One key issue is how much gas is displaced in the halo by the momentum 
of the ejecta after its energy is lost to cooling processes in the expanding flow at early times.  
In \ion{H}{2} regions the shock later strongly interacts with the dense shell of ionized gas expanding 
within the \ion{H}{2} region.  How well do metals in the remnant mix with the dense shell, and if 
dynamical instabilities arise that fragment the shock into enriched clumps, are they liable to 
collapse into new stars?  Numerical simulations excluding these phenomenon indicate that metals do 
not significantly mix with the IGM until gas falls back into the dark matter potential after about 
a Hubble time \citep{get07}.

Capturing the true evolution of the blast presents numerical challenges due to the spatial 
scales at play.  To accurately represent the initial ejecta as a free expansion, its 
hydrodynamical profile cannot extend beyond radii that enclose more surrounding gas than ejecta 
mass.  In the high central densities of neutral halos such radii are 5 $\times$ 10$^{-4}$ - 1 
$\times$ 10$^{-3}$ pc, but the blast must then be evolved to tens or hundreds of parsecs to evaluate 
its true impact on the halo.  Furthermore, multiple sites of intense luminosity will erupt in 
the flow over all these scales from the earliest times, so the chemistry of the flow must be 
solved consistently with hydrodynamics to determine where gas is collisionally ionized and 
therefore strongly radiating.  The attendant time scales can restrict integration times to much 
lower values than Courant times. 

To address these questions we have performed one-dimensional models of Type II supernovae, 
hypernovae, and PISN in neutral and ionized cosmological halos at redshift 20.  Our initial 
conditions are spherically-averaged halos derived from cosmological initial conditions with 
the Enzo adaptive mesh refinement (AMR) code \citep{enzom} that are photoionized by the star 
in the ZEUS-MP reactive flow radiative transfer code \citep{wn06}.  The aim of our survey is 
to determine the explosion energies required to destroy cosmological halos that are thought 
to form primordial stars by H$_2$ cooling.  We also evaluate their observational signatures 
and how their metals will mix with the ambient medium as a prelude to three-dimensional AMR
calculations now in progress.

Our radiation hydrodynamics scheme and initial profiles for the free expansion are presented 
in $\S$ 2.  We also describe the expanding grids used in this study and the time-dependent 
boundary conditions that present the ambient medium to the expanding flow.  \ion{H}{2} region profiles 
for the explosions and the grid of supernova and halo models selected for this survey are outlined 
in $\S$ 3.  The evolution of blasts in both neutral halos and \ion{H}{2} regions are examined together 
with dominant energy loss mechanisms for each case in $\S$ 4.  We discuss observational measures 
for both kinds of explosion in $\S$ 5, chemical enrichment of baryons in the halos in $\S$ 6 and 
conclude in $\S$ 7.

\section{Numerical Method} 

Three-dimensional halo baryon profiles computed from cosmological initial conditions are 
extracted from an Enzo AMR run and spherically averaged into one-dimensional density fields.
These profiles are then photoionized over the main sequence lifetime of the SN progenitor
with the ZEUS-MP astrophysical fluid hydrodynamics code.  We then initialize one-dimensional 
explosion profiles in the ionized halos and evolve the blasts in a separate calculation.  
ZEUS-MP has been modified to solve explicit finite-difference approximations to Euler's 
equations of fluid dynamics coupled to 9-species primordial chemistry with photoionization 
rate coefficients computed from multifrequency photon-conserving UV transport \citep{wn07b}.  
The fluid equations are  
\vspace{0.1in}
\begin{eqnarray}
\frac{\partial \rho}{\partial t}  & = & - \nabla \: \cdotp \; (\rho {\bf v})  \\
\frac{\partial \rho v_{i}}{\partial t}  & = & - \nabla \: \cdotp \; (\rho v_{i} 
{\bf v}) \: - \: \nabla p \: - \: \rho \nabla \Phi \: - \: \nabla \cdotp {\bf Q}    \\ 
\frac{\partial e}{\partial t}  & = & - \nabla \: \cdotp \; (e {\bf v}) \: - \: p\nabla \: 
\cdotp \: {\bf v} \: - \: \bf{Q} : \nabla  {\bf v} 
\end{eqnarray} \vspace{0.05in} 

\noindent 
where $\rho$, e, and the v$_{i}$ are the mass density, internal energy density, and velocity 
components of each zone, and p = ($\gamma$-1) e and {\bf{Q}} are the gas pressure and the von 
Neumann-Richtmeyer artificial viscosity tensor \citep{sn92}.  

Nine additional continuity equations are solved to advect H, H$^+$, He, He$^+$, 
He$^{++}$, H$^-$, H$^{+}_{2}$, H$_{2}$, and e$^-$, which are assumed to share a common 
velocity distribution.  Thirty reactions among these primordial species are followed with 
the nonequilibrium rate equations of \citet{anet97} 
\vspace{0.05in}
\begin{equation}
\frac{\partial \rho_{i}}{\partial t} = \sum_{j}\sum_{k} {\beta}_{jk}(T){\rho}_{j}{\rho}_{k} + \sum_{j} {\kappa}
_{j}{\rho}_{j} \vspace{0.05in}
\end{equation}
where ${\beta}_{jk}$ is the rate coefficient of the reaction between species j and k that 
creates (+) or removes (-) species i, and the ${\kappa}_{j}$ are the ionization rates.  The 
continuity equations for each reactant are updated in the advection routines and the 
reaction network is evolved in a separate operator-split semi-implicit substep.  We enforce 
charge and baryon conservation at the end of each hydrodynamic cycle by assigning any error 
between the species or charge sums and $\rho$ to the largest of the species to bring them 
into agreement with $\rho$.  Microphysical cooling and heating are included with an 
isochoric operator-split update to the energy density evaluated each time the reaction 
network is solved:  \vspace{0.05in}
\begin{equation}
{\dot{e}}_{\mathrm{gas}} = \Gamma - \Lambda \label{eqn: egas}
\vspace{0.05in}  
\end{equation}
Here, $\Gamma$ is the rate at which photons at all frequencies in the calculation deposit 
heat into the gas as described in \citet{wn07b} and $\Lambda$ is the sum of the cooling 
rates due to collisional ionization and excitation of H and He, recombinations in H and 
He, inverse Compton scattering (IC) from the CMB, and bremsstrahlung emission.  H$_2$ cooling
\citep{gp98} is included in the \ion{H}{2} region calculation but not in the evolution of the
SN blast, whose high temperatures and strong shocks destroy the fragile hydrogen molecules
on the time scales of our models.  We do not include metal line cooling, but reserve it
for three-dimensional models now in preparation.

Photon-conserving UV transport in the spherical polar coordinate grid used in this study
is performed by solving the static approximation to the equation of radiative transfer
in flux form 
\begin{equation}
\nabla \: \cdotp \; {\bf F} \: = \: - \chi\, {\bf F}_r,
\end{equation}
\noindent 
where $\chi$ is the inverse mean free path of a UV photon in the neutral gas
\begin{equation}
 \chi = \displaystyle\frac{1}{n \sigma},
\end{equation}
to compute the total absorption rate in each zone by photon conservation, which 
mandates that the number of absorptions in a zone is equal to the photons entering
the zone minus those exiting each second.  In terms of flux, this difference reduces 
to \vspace{0.1in}
\begin{equation}
\begin{array}{rcl}
n_{\mathrm{abs}} & = &\,F_{i}\, \left( 1 - e^{{-\chi \left(r_{i+1}-r_{i}\right)}}\right)\ \\
&&\\
          & \times & \:  \displaystyle\frac{r_{i}^{2}\, \left({\phi}_{k+1} - {\phi}_{k}\right)\left(\cos {\theta}_{j} - 
\cos {\theta}_{j+1} \right)}{h\nu}
\end{array}
\end{equation}
Individual interaction rates (photoionizations, photodetachments, and 
photodissociations) are obtained from the total rate according to the 
prescription \vspace{0.1in}
\begin{equation}
n_{i} = \displaystyle \frac{1 \: - \: e^{- \tau_i}}{\sum_{i=1}^{n}1 \: - \: 
e^{- \tau_i}} \, n_{\mathrm{abs}}. \vspace{0.075in}
\end{equation}   
which are then converted into the photoionization rate coefficients required by the
reaction network  \vspace{0.15in}
\begin{equation}
k_{i} = n_{i} \cdot \displaystyle \frac{1}{n_{\mathrm{spec}}\,V_{\mathrm{cell}}},\vspace{0.1in}
\end{equation}
where n$_{spec}$ is the number density of the species with which the photons 
are interacting.  We adopt the on-the-spot (OTS) approximation by assuming that 
recombinations to the ground state balance photoionizations by diffuse photons 
within a zone, so we do not explicitly transport recombination radiation.  Complete 
descriptions of the photon conserving UV transport from which radiative rates in the 
reaction network are derived can be found in \citet{wn06,wn07b}.

\subsection{Free-Expansion Blast Profiles} 

\label{subsec:SNprofile}

We take density and velocity profiles for the free expansion from \citet{tm99}, where it is 
shown that at a given time the ejecta density can be approximated by a flat inner core and 
steeply declining outer edge:

\begin{equation} \label{eq:SNdens}
\rho(v,t) = \left\{ 
\begin{array}{lcr}
F \cdot t^{-3} & \mbox{for} & v \leq v_{\mathrm{core}} \\
F \cdot t^{-3} \cdot (\frac{v} {v_{\mathrm{core}}}) ^ {-n} & \mbox{for} & v_{\mathrm{core}} < v \leq
v_{\mathrm{max}} \\
0. & \mbox{for} & v > v_{\mathrm{max}}
\end{array}
\right. ~,
\end{equation}

\begin{equation} \label{eq:freeexp}
v(r,t) = \frac{r}{t}~~\mathrm{for}~~t > 0 ~.
\end{equation}

\noindent
Here, $\rho$ is the density, $t$ is the time, $v$ is the velocity and $r$ is the radius. $F$ and 
$v_{\mathrm{core}}$ are normalization constants which must be determined. The time dependence in 
the density is due to the expansion of the ejecta: as it grows in volume by $r^3$ its density 
decreases by $r^{-3} \propto t^{-3}$.  The inner region of the free expansion, $v \leq v_{\mathrm{core}}$, 
has a constant density and the outer region, $v > v_{\mathrm{core}}$, decreases by a power law 
with index $n$. Such profiles, with $n =$ 9, are usually assumed for core collapse supernovae 
\citep{tm99,dw05,cc06}.  The early evolution of the blast is insensitive to the choice of 
$r_{\mathrm{max}}$, unless it exceeds the radius enclosing an ambient mass greater than the ejecta 
mass.   

The two normalization constants, $F$ and $v_{\mathrm{core}}$, are determined from the ejecta mass 
$M_{\mathrm{ej}}$ and energy $E_{\mathrm{ej}}$.  The energy is assumed to be entirely kinetic. To 
calculate these constants we first choose a maximum radius $r_{\mathrm{max}}$ ($> 0$) that defines 
the leading edge of the SN ejecta and a maximum velocity $v_{\mathrm{max}}$ for the remnant.  This 
radius is a free parameter and is usually set so that the enclosed baryonic mass in the halo equals
$M_{\mathrm{ej}}$.  The maximum velocity is set to $3 \times 10^4$ km/s, corresponding roughly to 
that observed in core-collapse supernova explosions. We set $r_{\mathrm{min}} =$ 0 in all our models.

To determine $F$ and $v_{\mathrm{core}}$, we first set $t = t_{\mathrm{max}} = r_{\mathrm{max}}/
v_{\mathrm{max}}$, the time by which the leading edge of free expansion has self-silimilarly grown
to $r_{\mathrm{max}}$, the initial radius we have chosen for the blast.  Using equations 
\ref{eq:SNdens} and \ref{eq:freeexp} and choosing $F$ = 1 and $v_{\mathrm{core}}$ to be some small
arbitrary value, we perform the following integrations:

\begin{equation} \label{eq:Mej}
M_{\mathrm{ej}} = \int_{r_{\mathrm{min}}}^{r_{\mathrm{max}}} \, 4\pi r^2 \cdot \rho(v,t) \mbox{d} r ~.
\end{equation}

\begin{equation} \label{eq:Eej}
E_{\mathrm{ej}} = \int_{r_{\mathrm{min}}}^{r_{\mathrm{max}}} \, 4\pi r^2 \cdot \frac{1}{2} \rho(v,t) v(t)^2 \mbox{d} r ~,
\end{equation}

\noindent
Neither of these integrations will return correct values for $M_{\mathrm{ej}}$ and $E_{\mathrm{ej}}$
at first because $F$ and $v_{\mathrm{core}}$ were arbitrarily chosen.  Selecting a new $F'$ equal 
to $M_{\mathrm{ej}}$ divided by the mass obtained from Eq. \ref{eq:Mej}, which guarantees the first 
integration equals the correct ejecta mass by construction, we repeat the second integration and 
check if it is equal to $E_{\mathrm{ej}}$.  If so, $F'$ and $v_{\mathrm{core}}$ are those needed 
to construct the density profile.  If not, we increase the value for $v_{\mathrm{core}}$, again set 
$F =$ 1, obtain a new $F'$, which partly depends on the new $v_{\mathrm{core}}$, from the first
integral, and then compute a new $E_{\mathrm{ej}}$ with the second integral.  We repeat this two-step 
procedure until the latest choice of $v_{\mathrm{core}}$ and $F'$ returns the correct value of $E_{\mathrm{ej}}$ 
in the second integration; the $F'$ in this final cycle is the $F$ we apply to the density profile 
on the grid.  Once we have $F$ and v$_{\mathrm{core}}$, and recalling that $t = t_{\mathrm{max}} = 
r_{\mathrm{max}}/v_{\mathrm{max}}$ and $v = r/t_{\mathrm{max}}$, we recast eq \ref{eq:SNdens} as a 
function of $r$ in order to initialize densities on the grid.

Analytical expressions can be found for the normalization constants $F$ and $v_{\mathrm{core}}$ 
without resorting to iteration in certain circumstances. Equations (\ref{eq:SNdens}) 
and (\ref{eq:freeexp}) can be substituted into Equations (\ref{eq:Eej}) and (\ref{eq:Mej}) 
to obtain

\begin{equation}
\label{eq:Mej2}
M_{\mathrm{ej}} = 4 \pi t_{\mathrm{max}}^{-3} F \left( \frac{r_{\mathrm{core}}^{3} - r_{\mathrm{min}}^{3}}{3} - r_{\mathrm{core}}^n\frac{(r_{\mathrm{max}}^{3-n} - r_{\mathrm{core}}^{3-n})}{3-n} \right) ~,
\end{equation}

\begin{equation}
\label{eq:Eej2}
E_{\mathrm{ej}} = 2 \pi t_{\mathrm{max}}^{-5} F \left( \frac{r_{\mathrm{core}}^{5} - r_{\mathrm{min}}^{5}}{5} - r_{\mathrm{core}}^n\frac{(r_{\mathrm{max}}^{5-n} - r_{\mathrm{core}}^{5-n})}{5-n} \right) ~,
\end{equation}

\noindent
Here, $t_{\mathrm{max}} = r_{\mathrm{max}}/v_{\mathrm{max}}$, $r_{\mathrm{core}} = v_{\mathrm{core}} \cdot t_{\mathrm{max}}$ and 
$r_{\mathrm{min}}$ and $r_{\mathrm{max}}$ are the minimum and maximum radius of the free expansion, 
respectively. If $r_{\mathrm{min}} \ll r_{\mathrm{core}} \ll r_{\mathrm{max}}$, which is generally true, then $F$ and 
$v_{\mathrm{core}}$ can be expressed as \citep{tm99}:

\vskip2mm
\begin{equation} \label{eq:v}
v_{\mathrm{core}} = \frac{r_0}{t_{\mathrm{core}}} = \left( \frac{10E_{\mathrm{ej}}(n-5)}{3M_{\mathrm{ej}}(n-3)} \right) ^ {\frac{1}{2}}  ~,
\end{equation}

\begin{equation} \label{eq:F}
F = \frac{1}{4\pi n} \cdot \frac{ (3(n-3)M_{\mathrm{ej}})^{\frac{5}{2}} } { (10(n-5)E_{\mathrm{ej}})^{\frac{3}{2}} } = \frac{10(n-5)E_{\mathrm{ej}}}{4\pi n} \cdot v_{\mathrm{core}}^{-5} ~.
\end{equation}
\vskip2mm

\noindent
We iterate to find $F$ and $v_{\mathrm{core}}$ because the assumptions underlying the analytical forms
might not be satisfied for all ejecta profiles discretized on a mesh. At times $r_{\mathrm{core}}$ could 
lie close to the inner or outer boundary of the grid. In this case the analytical solution is 
different from the one found iteratively and could introduce errors in the mass and energy 
on the grid if applied.  To prevent such errors we always solve for $F$ and $v_{\mathrm{core}}$ 
numerically.

\subsection{Moving Grid/Boundary Conditions}

As noted earlier, to realistically represent the early remnant as a free expansion in neutral halos it
cannot enclose more halo gas than its own mass in ejecta, which confines its initial extent to less
than 0.001 pc if the halo.  This restriction is problematic, given that its expansion must be accurately resolved
out to the virial radius of the halo, $\sim$ 100 pc.  However, this can be accomplished with a fixed 
number of mesh points if the grid itself expands with the flow and always maintains its outer boundary
beyond the expanding shock front.  In this scenario only the surrounding halo just beyond the shock
can be present on the grid at any time, so as the outer boundary of the grid grows with time
its densities, energy, and velocities must be updated according to its given position within the 
halo.  In this way the expanding flow always encounters the densities it would if the entire halo
resided on the grid.  An additional benefit is that Courant times become longer as the grid grows,
accelerating the execution of the calculation.

We activate moving grids in the radial coordinate in ZEUS-MP for the models in this study, modifying
the outer boundary conditions every hydrodynamical time step with an operator-split update following
the calculation of a new grid.  At the start of the simulation we allocate 80\% of the mesh to the 
free expansion profile, reserving the outer 20\% for the ambient halo. Every time step thereafter the 
outer 10\% of the grid is swept for the maximum radial gas velocity and its position on the mesh.  A 
grid velocity describing homologous expansion is then assigned to each grid point such that the grid 
velocity at the radius of 
the previous maximum in gas velocity is three times this maximum.  The factor of 3 guarantees that 
the flow never reaches the outer boundary, where inflow conditions are reset each time step to ensure 
the remnant always encounters the correct ambient medium.  If the velocity of the grid is 
non-zero, the entire radial grid is updated in the following manner. First a new outer boundary is determined from the old 
boundary, the grid velocity and the current time step.  A new grid is then computed such that the 
radial cells, whose number remains fixed, are redistributed equally over the new range in radius.
See \citet{het06} for a discussion of ZEUS-MP's moving Eulerian grid option. 

Species densities, energies, and velocities on the outer boundary are updated each hydrodynamical 
time step by interpolating from a data table whose entries are binned by radius.  The profiles in 
these tables are those of the halo photoionized by the original star over its main sequence lifetime, 
which were computed in the separate calculation with nine-species primordial chemistry and radiative 
transfer that was described above.  The flow values assigned to the outer boundary depend upon the 
position bin into which the boundary falls at a given time step.  We describe the stellar masses and 
halos used in our models in the following section.

\subsection{Static Dark Matter Potential}

Primordial supernova remnants evolved in much deeper gravitational potentials due to dark
matter than SN in the galaxy today.  Gravitational potentials in the halos in our study are 
dominated by baryons at radii of less than a parsec and by dark matter at larger radii.  We neglect 
the self-gravity of the remnant on scales below a parsec because of its large kinetic energy but 
include the potential of the dark matter on all scales.  Since merger times at $z \sim$ 20 are 
approximately 20 Myr, much longer than the dynamical times of the remnant, to good approximation 
the gravitational potential of the dark matter can be held fixed in our simulations. We interpolate 
the dark matter potential of the halo onto the initial grid at the beginning of the run and onto
each new grid thereafter, but the potential itself never evolves.  We apply this potential to force 
updates to the gas velocities throughout the calculation.  The interpolation substep is operator 
split from the time-dependent boundary updates of the fluid variables described above.  

\section{SN Progenitor \ion{H}{2} Regions}

\begin{deluxetable}{ccccc}
\tabletypesize{\scriptsize}
\tablecaption{Halos \label{tbl-1}}
\tablehead{
\colhead{halo} & \colhead{$J_{\mathrm{LW}}$} & \colhead{$M$ ($\Ms$)} & \colhead{$z_{\mathrm{coll}}$} &
\colhead{$E_{\mathrm{B}}$ (erg)}}
\startdata
 1 &     0    &  6.9E+05 & 24.1 & 1.53E+49\\
 2 & 3.16E-23 &  2.1E+06 & 20.4 & 1.10E+50\\
 3 & 1.00E-21 &  1.2E+07 & 17.3 & 2.50E+51\\
\enddata
\end{deluxetable}

\begin{figure}
\resizebox{3.45in}{!}{\includegraphics{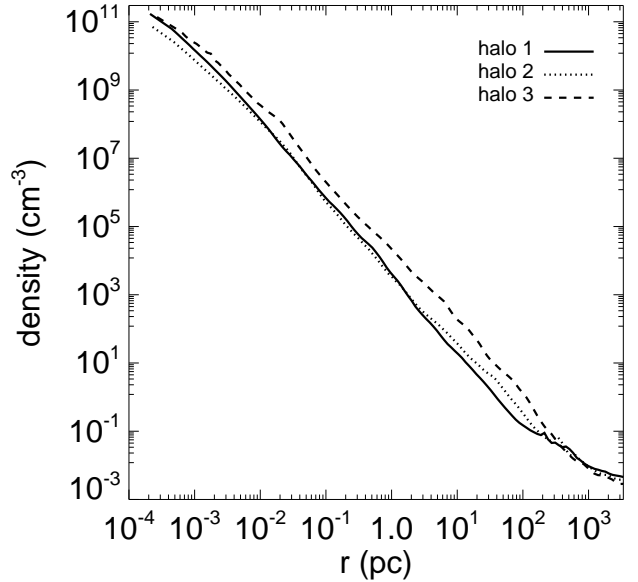}}
\caption{Spherically-averaged baryon density profiles for the halos examined in this study.} 
\label{halos}
\end{figure}

The halos adopted for our study were taken from \citet{oshea07a}, who examined the formation of
Pop III stars in a range of Lyman-Werner (LW) backgrounds (11.18 - 13.6 eV).  Since more baryons are 
required to shield H$_2$-cooled cores from dissociation as the LW background rises, the minimum halo 
mass required to form a star rises with decreasing redshift.  We consider three different halos that 
span the masses for which Pop III stars are expected to form by H$_2$ cooling.  Their properties are 
summarized in Table \ref{tbl-1} and their spherically-averaged baryon density profiles are shown in Figure
\ref{halos}.  In column 1, we list the halo number, in column 2 the LW background in which they form, 
in column 3 their mass, in column 4 their redshift, and in column 5 their binding energy.  SN 
progenitors were chosen according to the types of explosions that were possible in 
primordial halos: 15 $\Ms$ (Type II supernova), 40 $\Ms$ (hypernovae), and 260 $\Ms$ (PISN). 
Their properties are tabulated in Table \ref{tbl-2} \citep{s02}. To obtain accurate circumstellar media for the
supernovae, each star was allowed to ionize each halo and then explode, for a total of nine models 
listed in Table \ref{tbl-3}. In Table \ref{tbl-1} $J_{\mathrm{LW}}$ is the mean intensity of LW photons 
in which the halos form, centered at 12.87 eV in units of erg$^{-1}$ cm$^{-2}$ Hz$^{-1}$ sr$^{-1}$ 
and $z_{\mathrm{coll}}$ is the redshift at which the star appears in the halo. The binding energy of the gas
$E_{\mathrm{B}}$ in each halo can be approximated by that of a homogeneous sphere, which is given by
\begin{equation}
E_{\mathrm{B}} = \frac{3}{5} \; \frac{GM_{\mathrm{h}}m_{\mathrm{b}}}{R_{\mathrm{vir}}},
\end{equation}
where $M_{\mathrm{h}}$ and $m_{\mathrm{b}}$ are the total halo and baryon masses, respectively, and $R_{\mathrm{vir}}$ is the virial 
radius of the halo.  In Table \ref{tbl-2} ${\dot{n}}_{\mathrm{ph}}$ is the total number of ionizing photons 
emitted per second. 

\begin{deluxetable}{ccccc}
\tabletypesize{\scriptsize}
\tablecaption{SN Progenitors \label{tbl-2}}
\tablehead{
\colhead{$M$ ($\Ms$)} & \colhead{$t_{\mathrm{MSL}}$ (Myr)} & \colhead{log $T_{\mathrm{eff}}$} & 
\colhead{log $L$/$\Ls$} & \colhead {${\dot{n}}_{\mathrm{ph}}$ (s$^{-1}$)}}
\startdata
 15 & 10.4  & 4.759 & 4.324 & 1.861E+48 \\
 40 & 3.864 & 4.900 & 5.420 & 2.469E+49 \\
260 & 2.106 & 5.004 & 6.721 & 3.272E+50 \\
\enddata
\end{deluxetable}

\begin{figure*}
\epsscale{1.17}
\plottwo{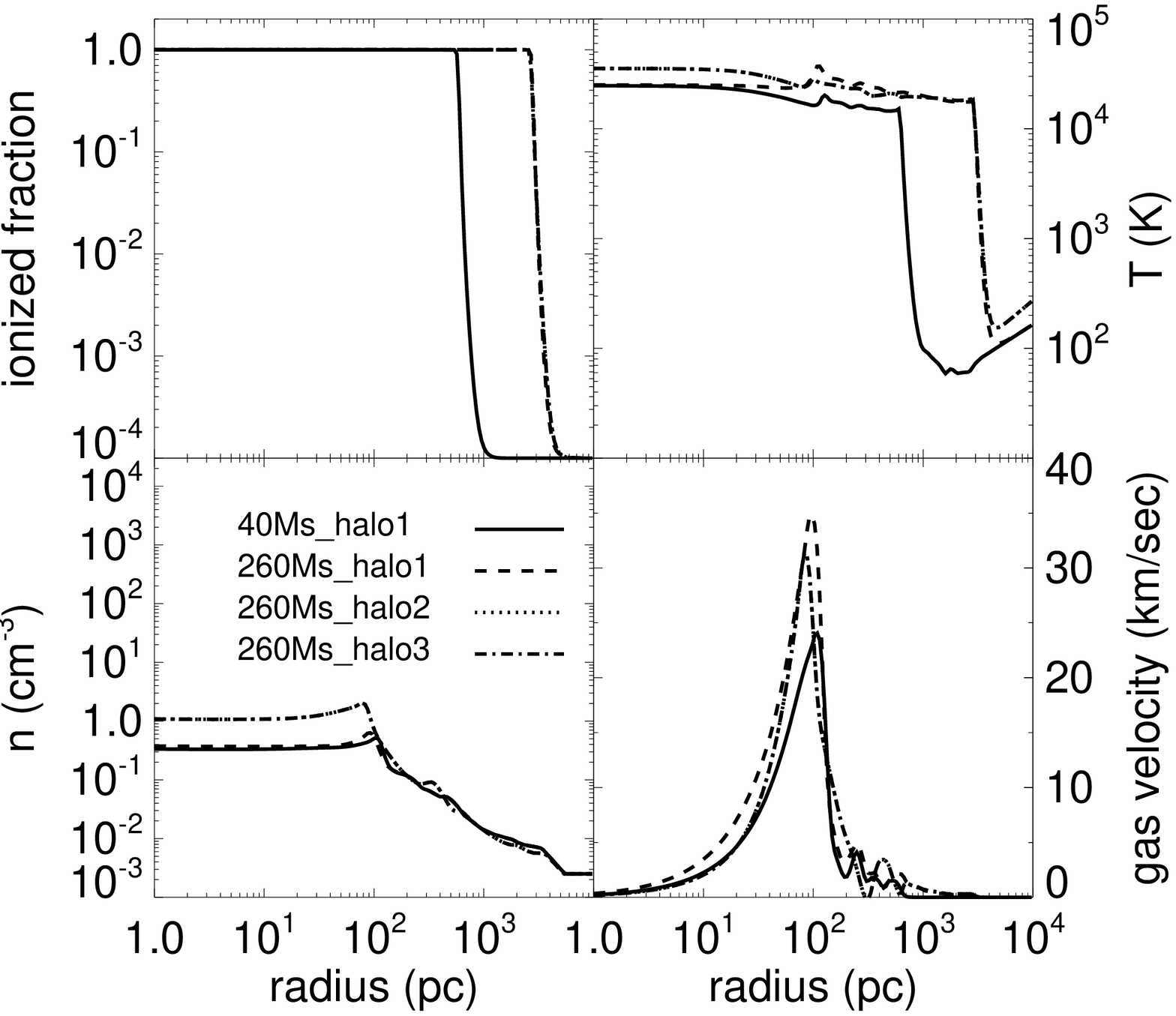}{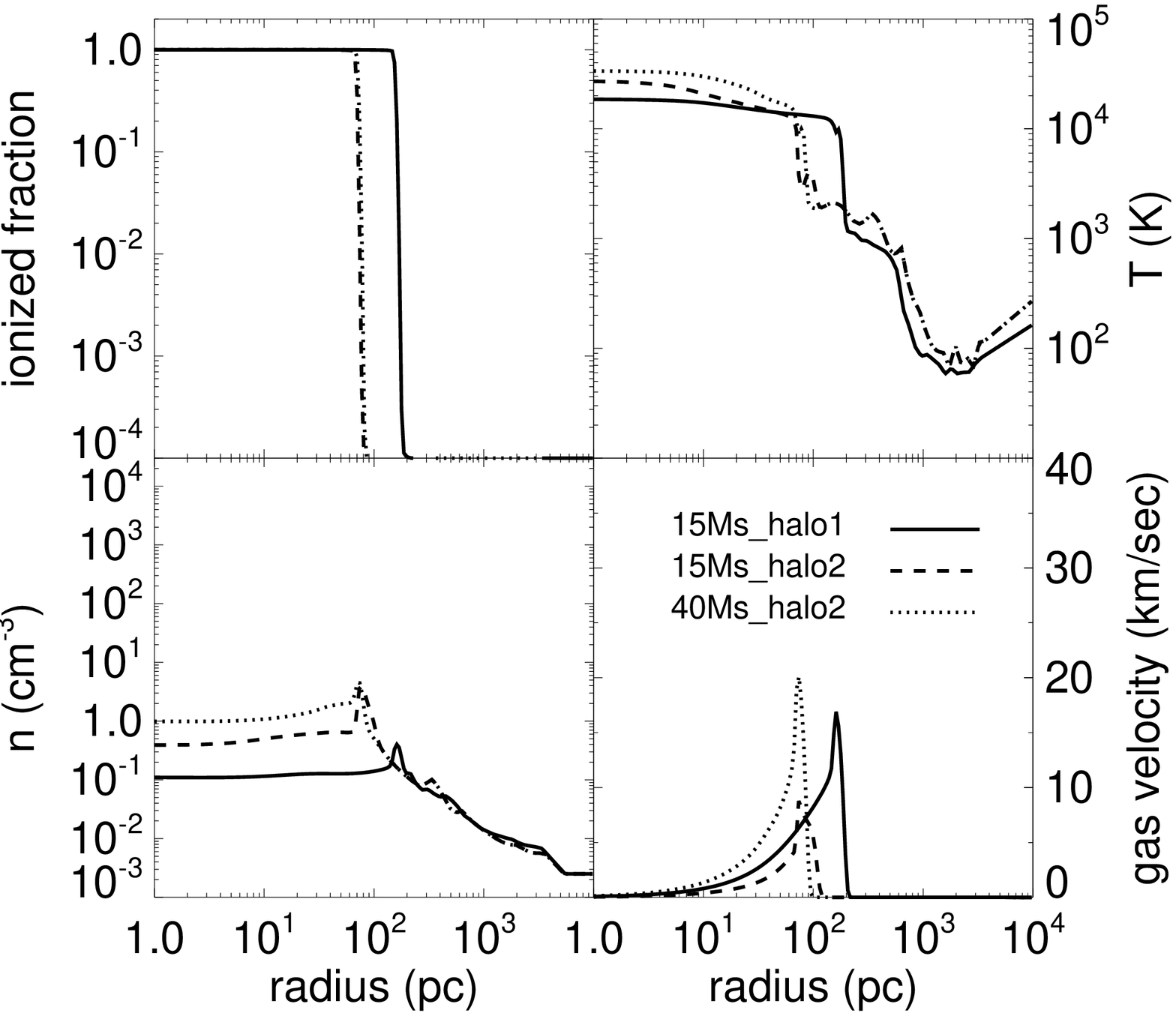}
\caption{Density, temperature, ionization fraction, and velocity profiles of the SN progenitor
\ion{H}{2} regions at the end of the life of the stars.  Left: \ion{H}{2} regions that break 
out of the halo.  Solid: 40Ms\_halo1; dashed:  260Ms\_halo1; dotted: 260Ms\_halo2; dot dash:
260Ms\_halo3.  Right: \ion{H}{2} regions that are confined to the virial radius of the halo.  
Solid: 15Ms\_halo1; dashed: 15Ms\_halo2; dotted: 40Ms\_halo2.  Since the \ion{H}{2} region
radii of the 15 $\Ms$ and 40 $\Ms$ stars in halo 3 are below the resolution limit of the 
radiation calculation, their hydrodynamical profiles are considered to be those of the 
neutral halo. \label{HIIregions}}  
\end{figure*}

\subsection{Problem Setup}

The star was centered in a 200 zone one-dimensional spherical coordinate mesh with reflecting
and outflow inner and outer boundary conditions, respectively.  Cell sizes were ratioed by the 
parameter $\beta$ according to the prescription
\begin{equation}
\displaystyle\frac{\Delta r_{i+1}}{\Delta r_{i}} = \beta. 
\end{equation}

\begin{deluxetable}{cccc}
\tabletypesize{\scriptsize}
\tablecaption{Primordial SN Explosion Models\label{tbl-3}}
\tablehead{
\colhead{halo} & \colhead{15 $\Ms$} & \colhead{40 $\Ms$} & \colhead{260 $\Ms$}}
\startdata
 1 & 15Ms\_halo1 & 40Ms\_halo1 & 260Ms\_halo1 \\
 2 & 15Ms\_halo2 & 40Ms\_halo2 & 260Ms\_halo2 \\
 3 & 15Ms\_halo3 & 40Ms\_halo3 & 260Ms\_halo3 \\
\enddata
\end{deluxetable}

We applied $\beta$ from 1.049 to 
1.055 to concentrate zones at the origin, both to resolve the central core and density 
drop and to obtain the maximum time step capable of accurately capturing the emergence of 
the I-front from the core.  The outer boundary in all the \ion{H}{2} region models was 10 kpc  
and the inner boundary of each model was chosen to enclose a volume at the center of the 
neutral halo equal in mass to the star.  In doing so, we assume that all of this enclosed 
gas goes into forming the star and that the densities just beyond this volume are those 
that the nascent I-front encounters. In lieu of more accurate models of the environment 
of the newborn star, these measures ensure the I-front breaks out into the proper ambient 
densities.  For a given star and halo, the inner boundary for the \ion{H}{2} region calculation
is thus different for the inner boundary chosen for the explosion, which is zero in our
models as noted in section 2.1.  This choice is inappropriate for the \ion{H}{2} region calculation
because it would present higher initial densities to the emerging front than it really 
encounters.  The inner problem radii vary from 2.6 $\times$ 10$^{-3}$ - 1.4 $\times$ 10$^{-2}$
pc.

Forty equally-spaced energy bins were allocated to the radiation transport from 0.755 eV to 
13.6 eV and 80 logarithmically-spaced bins were partitioned from 13.6 eV to 90 eV.  We found 
it necessary to include all nine species in our UV breakout models.  Breakout of the I-front 
from the 15 $\Ms$ star is borderline in the more massive halos and may be prevented if 
a dense shell forms in the front by H$_2$ cooling.  However, H$_2$ chemistry is problematic 
at the inner boundary as photons propagate through the central zone.  The hard UV spectrum 
significantly broadens the emergent front, resulting in a protracted period of partial 
ionization at warm temperatures in the first zone.  This, together with very large central
densities, results in unrealistically high H$_2$ formation rates that violate species 
conservation.  This difficulty was eliminating by simply turning off H$_2$ chemistry until
the central zone was completely ionized. Each model was evolved for the main sequence lifetime 
of the star in the halo.  We tabulate final \ion{H}{2} region radii for each model in Table 
\ref{tbl-4}.  Final densities, temperatures, ionization fractions, and velocities for all 
runs in which \ion{H}{2} regions form are shown in Figure \ref{HIIregions}.

\begin{deluxetable}{crrr}
\tabletypesize{\scriptsize}
\tablecaption{\ion{H}{2} Region Radii (pc)\label{tbl-4}}
\tablehead{
\colhead{halo} & \colhead{15 $\Ms$} & \colhead{40 $\Ms$} & \colhead{260 $\Ms$}}
\startdata
 1 & 161.5  & 599.2  & 2843  \\
 2 &  70.1  &  74.1  & 2694  \\
 3 & -------- & -------- & 2694  \\
\enddata
\end{deluxetable}

\subsection{\ion{H}{2} Region Breakout}

The \ion{H}{2} regions in this survey fall into two classes. Those confined to the virial radius 
of the halo or less and are bounded by subsonic D-type fronts are shown in the right panel
of Figure \ref{HIIregions}.  Those that flash-ionize the halo on relatively short time scales 
and propagate into the IGM as supersonic R-type fronts appear in the left panel of Fig
\ref{HIIregions}. The I-fronts of very massive primordial stars in cosmological halos first
transform from R-type to D-type at radii of less than a parsec but then revert back to R-type 
in a few hundred kyr and completely overrun the halo \citep{wan04,ket04}.  The I-front exits
so rapidly that steep drops in the halo density profile are frozen in place as they are
ionized.  Once ionized and isothermal, the density gradients become large pressure gradients 
that launch shocked flows outward from the center. Ionized core shocks can evict more than half 
of the baryons from the virial radius of the halo over the lifetime of the star, piling the gas 
up into a dense shell that surrounds an evacuated cavity of very diffuse gas ($\lesssim$ 0.1 
cm$^{-3}$).  This is evident for the 260 $\Ms$ star in halos 1, 2 and 3 in the left panel of 
Figure \ref{HIIregions}, which are ionized out to kpc scales. The velocity of the ionized core shock 
is determined both by the postfront temperature and the slope of the density profile when ionized: 
it is typically 20 - 40 km s$^{-1}$.  We note that the 40Ms\_halo1 model is an intermediate case in 
that the R-type front breaks through the neutral shell just at the end of the star's life and 
barely overruns the halo.

\subsection{Trapped \ion{H}{2} Regions}

Less massive stars drive weaker I-fronts that either do not revert to R-type or even propagate 
into the halo at all, as in the case of the 15 $\Ms$ or 40 $\Ms$ stars in halo 3.  At the 
inner radii at which we initialize the respective H II region calculations (2.0e-3 - 0.01 
pc), the densities in halo 3 are 2 - 5 times greater than those in halo 1, so recombination 
rates are greater there by a factor of 4 - 25.  There is also more pressure downward on the 
\ion{H}{2} region bubble in halo 3 because there is more overlying matter in its deeper 
potential well.  The lower photon emission rates of the 15 $\Ms$ and 40 $\Ms$ stars cannot 
compete with these effects, and the \ion{H}{2} region radius remains below the resolution 
limit of the radiation calculation.  Not even an ultracompact (UC) \ion{H}{2} region forms, 
so we take the halo to be essentially neutral and undisturbed at the end of the life of the 
star.  In contrast, the I-fronts in the 15Ms\_halo1, 15Ms\_halo2, and 40Ms\_halo2  
models do escape the core but cannot escape the virial radius.  They remain D-type, as seen in 
the relative positions of the breaks in the ionization fractions and the shocks in the right 
panel of Figure \ref{HIIregions}.  The density structure of the D-type fronts differs from the 
R-type fronts above. Since the expansion rate of the dense neutral is subsonic with respect to 
the sound speed in the \ion{H}{2} region, acoustic waves level the densities in the ionized 
gas to flat uniform values as the shock expands. 

We expect a supernova remnant's greatest energy losses in \ion{H}{2} regions to occur upon impact 
with the shocked neutral or ionized shell.  Prior to its encounter with the shock the remnant 
will generally experience greater radiative cooling in \ion{H}{2} regions bounded by D-type fronts 
than in halos that are flash-ionized because their interiors have somewhat higher densities.  
However, to first order the effect of any \ion{H}{2} region that forms in a cosmological halo is to
greatly reduce the ambient densities into which the SN ejecta erupt, from $\sim$ 10$^{11}$ 
cm$^{-3}$ to $\lesssim$ 1 cm$^{-3}$.  This prolongs the free-expansion of the blast and delays 
cooling.  Radiative and losses in neutral halos will be prompt because the remnant initially 
encounters much higher densities than in \ion{H}{2} regions.  In the majority of our models, the H 
II region of the star drives most of the baryons from the halo by photoevaporation, but a 
supernova explosion is still necessary to remove 90\% of the gas from the halo, which is the 
criterion for its destruction \citep{ky05}.  The final density, energy, velocity, and abundance 
profiles for each \ion{H}{2} region are tabulated in data files.  The time-dependent outer boundary 
updates interpolate these data to ensure that the SN remnant expands into the correct ambient 
medium.

\section{SN Blast Wave Evolution}  

\begin{figure}
\resizebox{3.45in}{!}{\includegraphics{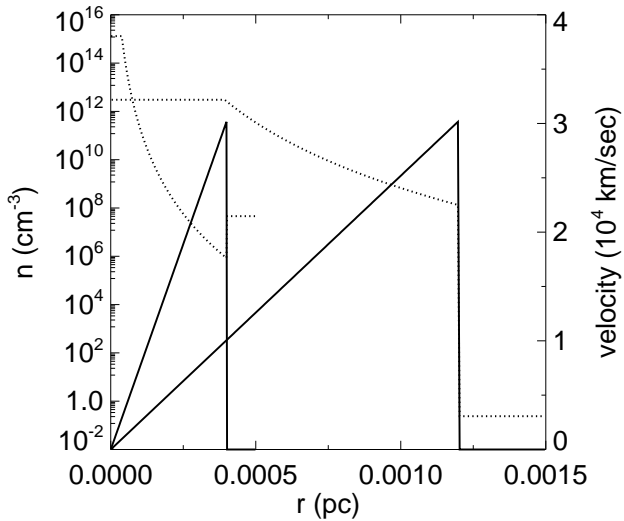}}
\caption{Density and velocity profiles of two free expansion blast profiles used in this
study.  Left: 15Ms\_halo3; right: 40Ms\_halo1.  Dotted: number density; solid: velocity.} 
\label{TM}
\end{figure}

The \citet{tm99} free-expansion solution for the supernova blast is completely parametrized
by the mass, kinetic energy, and maximum velocity of the ejecta, which are listed in Table
\ref{tbl-5} \citep{hw02,tun07}.  We also include the mass of the heavy elements expelled by
each supernova, $M_{\mathrm{el}}$, in Table \ref{tbl-5}.  It should be noted that 
pair-instability blast energies are not well constrained in present stellar evolution models, 
and that while the star is thought to be completely dispersed it is not clear how much of its 
mass goes into the ejecta.  For simplicity, we assume that all the mass of the PISN progenitor 
is in the free expansion.  We overlay initial density and velocity profiles for the 15Ms\_halo3 
and 40Ms\_halo1 models in Figure \ref{TM}.  Note the enormous disparity in ambient density between 
blasts initialized in neutral halos and in \ion{H}{2} regions.  We adopt a power law $n =$ 9 
in eq \ref{eq:SNdens} for all of our blast profiles.  The free expansion is initialized on a
spherical coordinate mesh with 250 uniform zones with reflecting and outflow boundary
conditions, respectively.  The free expansion occupies the first 200 zones and is joined 
directly without smoothing or interpolation to the circumstellar medium, which occupies
the outer 50 zones.  The temperature of the ejecta is set to 1000 K (the temperature of
the ambient medium is determined by the gas energy density interpolated in each zone from
the halo table data).  Since it is relatively cool, the ejecta is assumed to be nearly
neutral, with only the small ionization fraction $\sim$ 10$^{-4}$ expected for primordial
gas at $z \sim $ 20.  In reality, whether the ejecta is ionized or neutral matters little
because the energy required for its complete ionization is only a small fraction of the 
energy of the blast and is not significant to its kinematics, as will be shown later. 

We deactivate the transport of ionizing radiation during the evolution of the ejecta but
retain nine-species primordial chemistry and gas energy updates due to radiative cooling
with the hydrodynamics.  Collisional excitation and ionization cooling of H, He, and He$^+$,
recombinational cooling of H$^+$, He$^+$, and He$^{++}$, bremsstrahlung cooling, and inverse Compton
cooling are included in the energy equation.  The inner boundary of each problem was $r =$ 
0 and the outer boundary was that which would enclose a volume of neutral halo equal in 
mass to the ejecta, whether or not the halo is ionized by the star.  For halos 1 and 3 the
outer problem radius was 0.0005 pc and for halo 2 it was 0.0015 pc, whether or not the halo
was ionized by the progenitor.  This guarantees that the blast has not yet swept up its own 
mass in the halo or departed from a free expansion, and that all energy losses from the 
remnant are captured from the earliest times.  Such small radii for the outer boundary are 
unnecessary for explosions in \ion{H}{2} regions, in which blasts remain free expansions 
for a parsec or more, but are adopted anyway for simplicity.

\begin{deluxetable}{ccccc}
\tabletypesize{\scriptsize}
\tablecaption{Blast Parameters \label{tbl-5}}
\tablehead{
\colhead{star} & \colhead{M$_{\mathrm{ej}}$ ($\Ms$)} & \colhead{M$_{\mathrm{el}}$ ($\Ms$)} & \colhead{$E_{\mathrm{ej}}$ ($\Ms$)} & \colhead{v$_{\mathrm{max}}$ (km s$^{-1}$)}}
\startdata
 15 $\Ms$  & 13.52 & 5.94   & 1.0E+51 &  30000 \\
 40 $\Ms$  & 34.43 & 20.43  & 3.0E+52 &  30000 \\
 260 $\Ms$ & 260   & 125.27 & 1.0E+53 &  30000 \\
\enddata
\end{deluxetable}

Primordial supernovae in cosmological halos follow two distinct evolutionary paths 
depending on whether they explode in neutral clouds or in \ion{H}{2} regions.  We now 
examine the evolution of each type of remnant.

\subsection{Supernovae in \ion{H}{2} Regions}

In Figures \ref{fig:260Ms_halo1_1} and \ref{fig:260Ms_halo1_2} we show four stages of a 260 
$\Ms$ pair-instability supernova in halo 1, which is completely ionized by the progenitor.  
We show in Figure \ref{fig:losses} cumulative radiation losses together with kinetic energy
for all nine models.  The radiative processes tallied are H, He, and He$^+$ collisional 
excitation and ionization losses, bremsstrahlung, and IC scattering from the CMB.

\subsubsection{$t <$ 10 yr: Initial Shock}

\begin{figure*}
\epsscale{1.17}
\plottwo{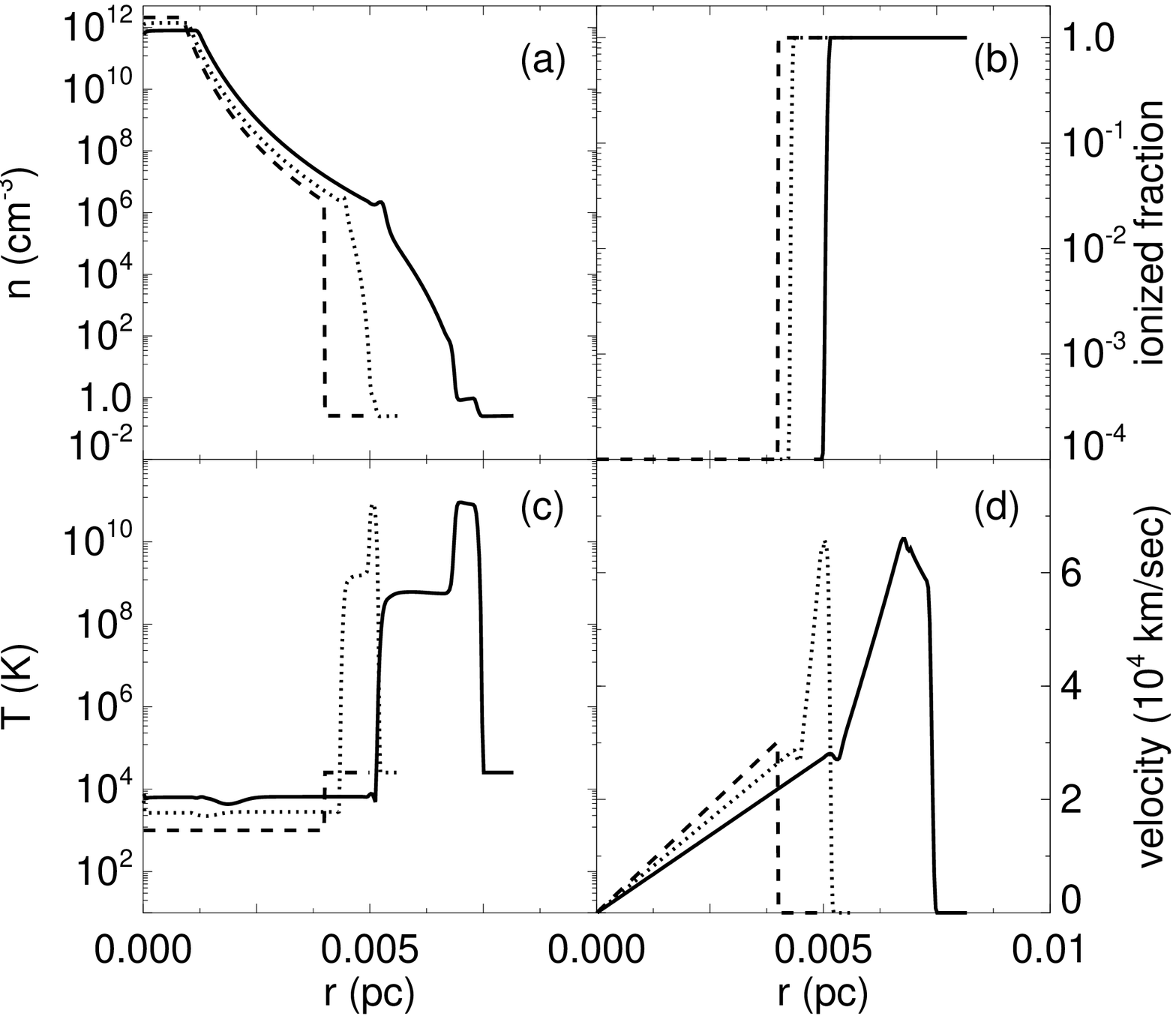}{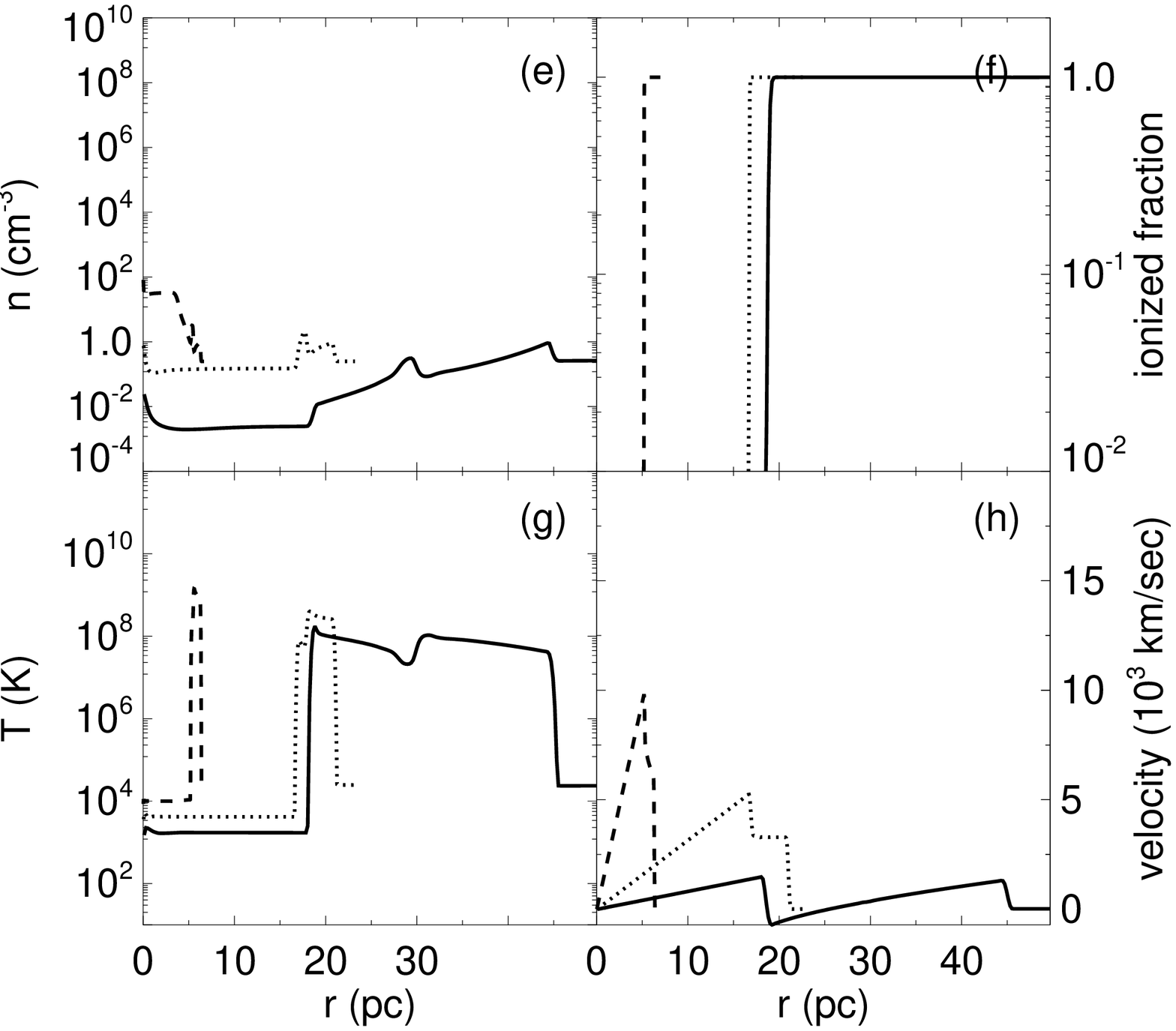}
\caption{Early evolution of the 260 $\Ms$ pair-instability supernova in halo 1.  
Panels (a) - (d): early free expansion. Dashed: $t = $ 0; dotted: 2.9 days; solid: 14.5 
days. Panels (e) - (h): end of the free expansion.  Dashed: $t = $ 515.6 yr; dotted: $t= 
$ 3093 yr; solid: $t= $ 6347 yr.
\label{fig:260Ms_halo1_1}}  
\end{figure*}

\begin{figure*}
\epsscale{1.17}
\plottwo{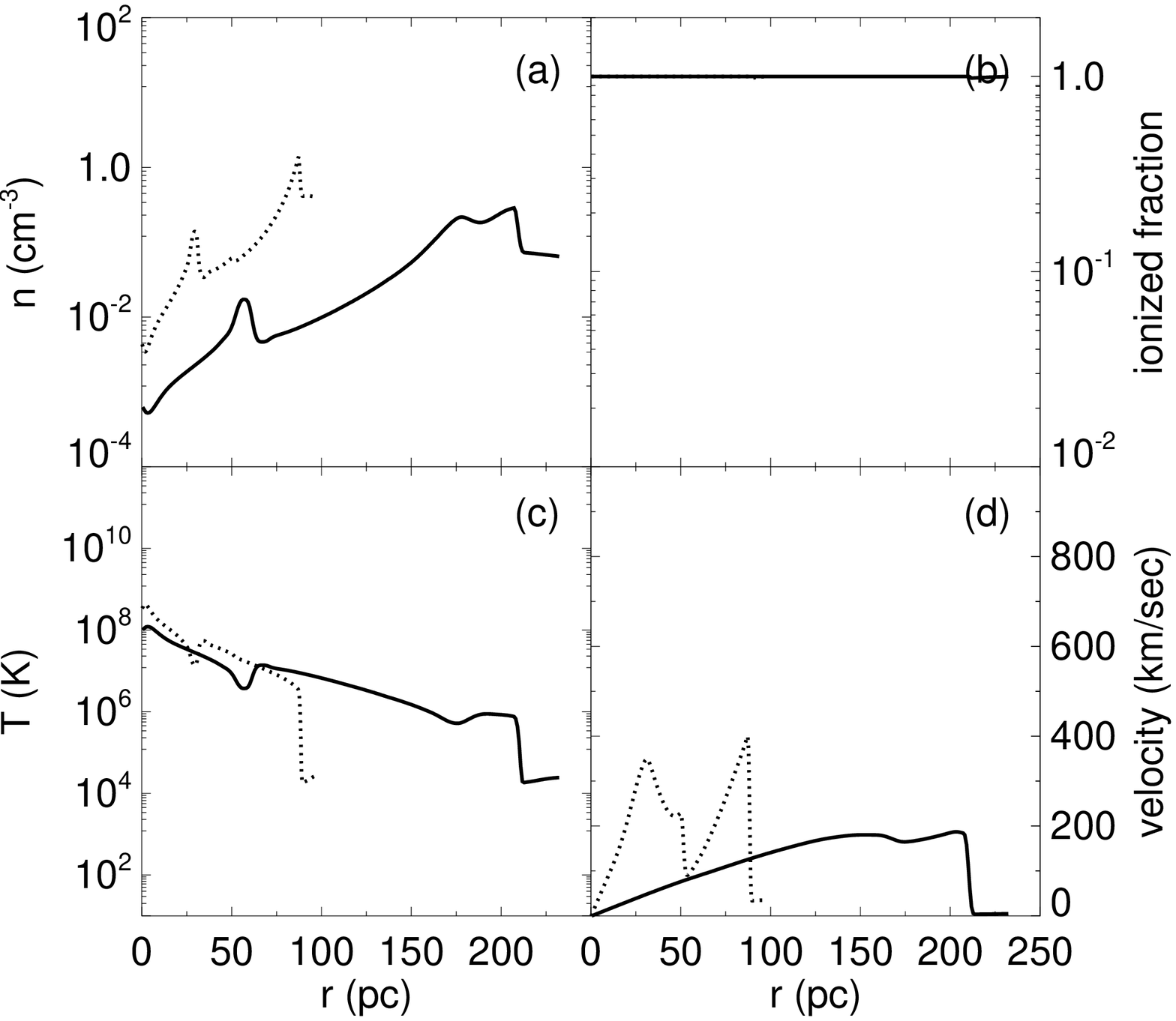}{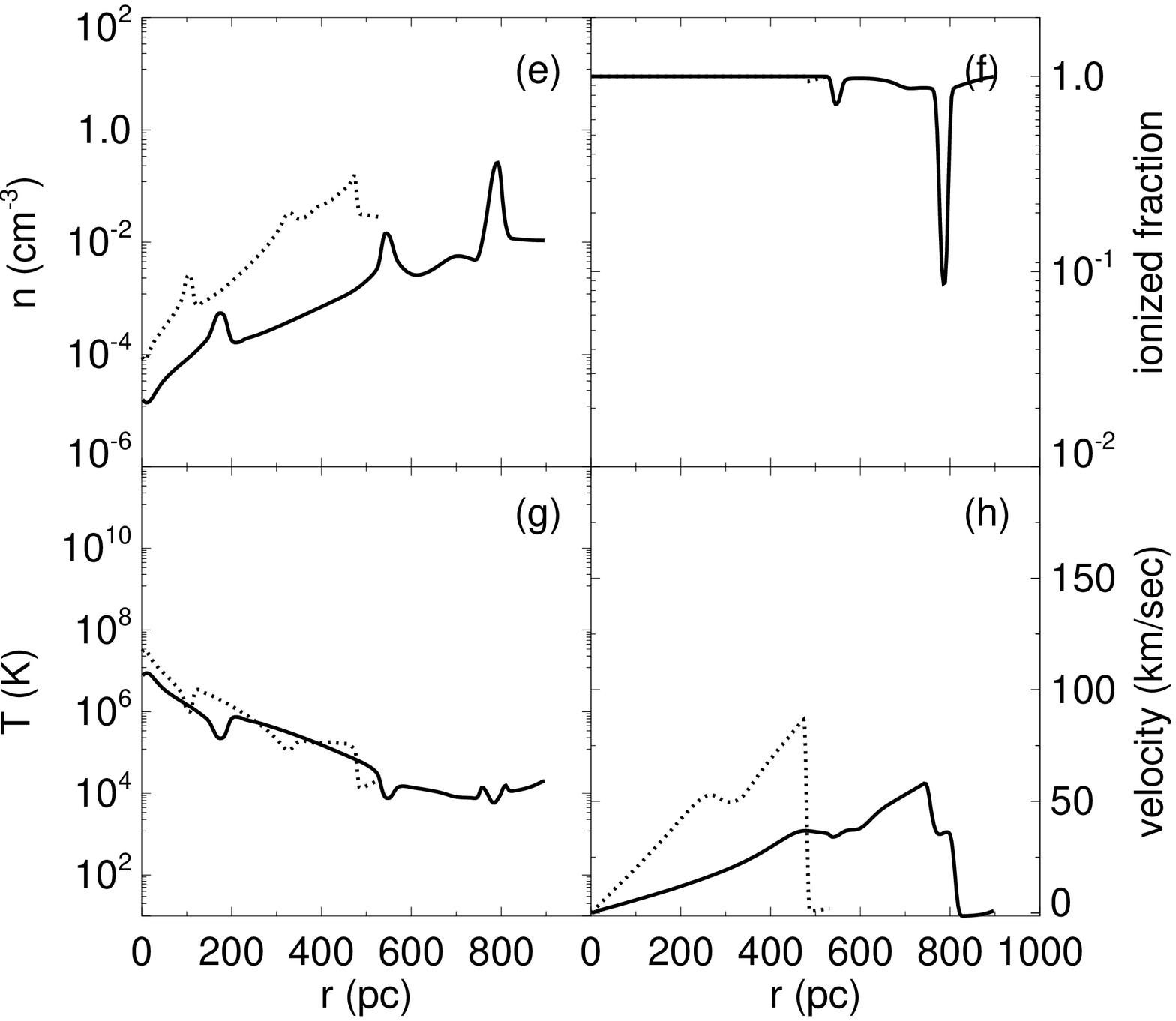}
\caption{Later evolution of the 260 $\Ms$ pair-instability supernova in halo 1.  
Panels (a) - (d):  interaction of the 260 $\Ms$ pair-instability remnant with 
the relic ionized core shock.  Dotted: $t = $ 19.8 kyr; solid: $t =$ 420 kyr, 
respectively.  Panels (e) - (h): flow profiles at 2.0 Myr (dotted) and 7.93 Myr 
(solid).
\label{fig:260Ms_halo1_2}}  
\end{figure*}

\begin{figure*}
\epsscale{1.15}
\plotone{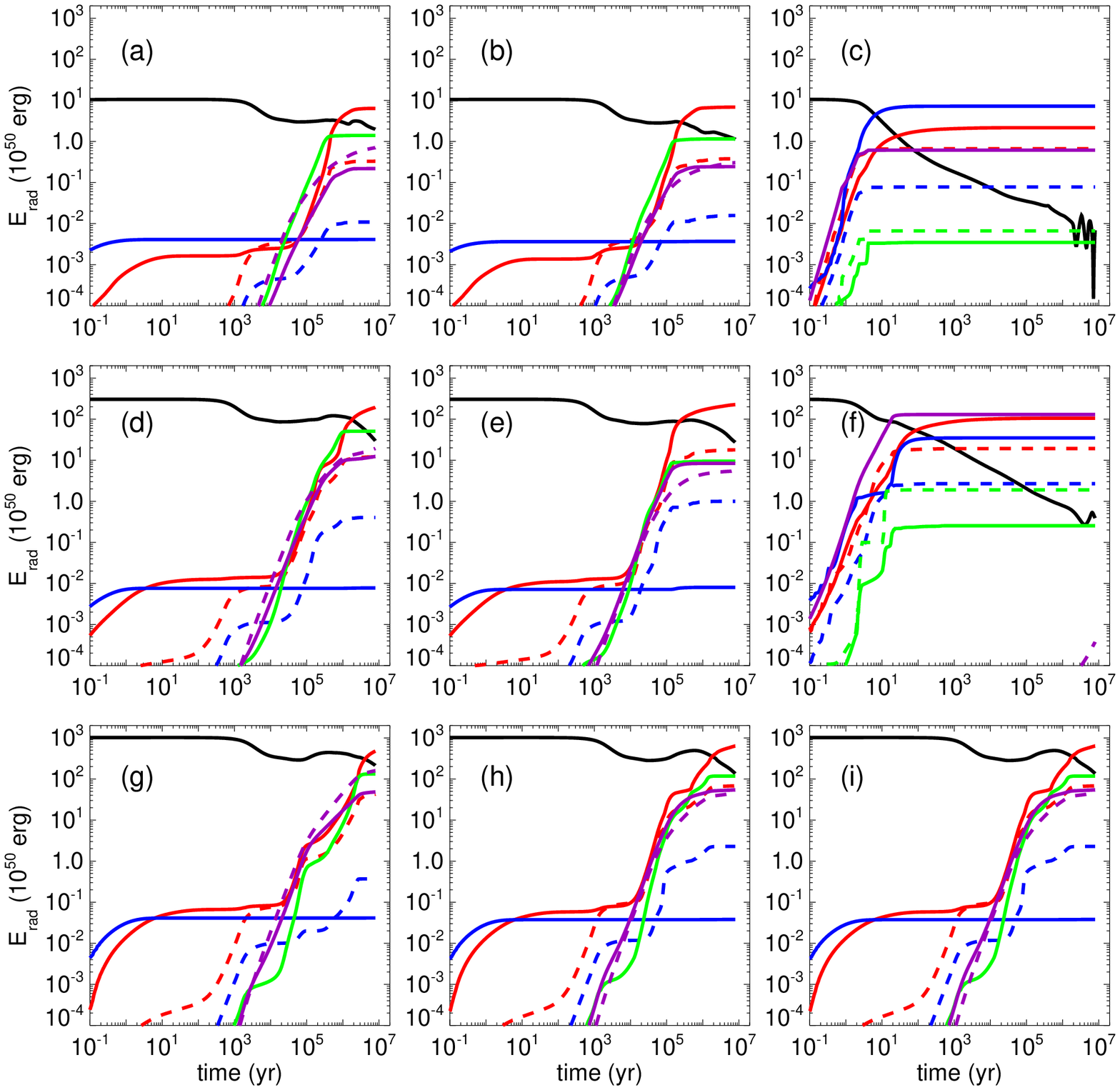}
\caption{Radiative energy losses in all nine remnants $vs$ time.  Black solid: total kinetic
energy; red solid: collisional excitation of H; red dashed: collisional ionization 
of H; blue solid: collisional excitation of He; blue dashed: collisional ionization 
of He; green solid: collisional excitation of He$^+$; green dashed: collisional 
ionization of He$^+$; purple solid: bremsstrahlung; purple dashed: inverse Compton 
scattering.  Panels (a) - (c):  15 solar mass Type II SN in halo 1, halo 2, and halo 3, 
respectively. Panels (d) - (f):  40 solar mass hypernova in halo 1, halo 2, and halo 3,
respectively.  Panels (g) - (h):  260 solar mass pair-instability SN in halo 1, halo 2, 
and halo 3, respectively.
\label{fig:losses}}  
\end{figure*}

Even though its surroundings are diffuse, n $\lesssim$ 0.1 cm$^{3}$, the free expansion 
drives a shock into the \ion{H}{2} region that heats immediately to over 10$^{11}$ K, 
as shown in Figure \ref{fig:260Ms_halo1_1}c.  The shocked gas is collisionally ionized and 
accelerates to 67000 km s$^{-1}$.  This occurs because the abrupt drop in density at 
$r \sim$ 0.004 pc in the initial blast profile generates a sharp pressure gradient when 
it is ionized, and hence isothermal.  The gradient accelerates the small amount of gas 
lying within it to nearly three times its original velocity, driving a rarefaction wave 
backward into the denser gas behind that is visible in the velocity profile at $r \sim$ 
0.0045 pc at $t =$ 2.9 days.  The drop in velocity of the dense gas in this wave and the 
speedup at the edge of the expansion are insignificant compared to the blast energy, so 
the kinetic energy of the ejecta remains steady for $t <$ 700 yr, as seen in the solid 
black plot in Figure \ref{fig:losses}g.  The temperature of the ionized gas in the 
rarefaction wave falls from 10$^{11}$ - 2 $\times$10$^{9}$ K at 2.9 days and 
then to 7 $\times$ 10$^{8}$ K by 14.5 days in exchange for $PdV$ work on its surroundings.  
We note that the density jump conditions for a strong shock are satisfied at $r \sim$ 
0.0075 pc at 14.5 days.

The early H and He excitational losses over $t <$ 10 yr in Figure \ref{fig:losses}g are 
due to the ionization of forward edge of the ejecta.  They are minute compared to the 
kinetic energy of the ejecta, which freely expands until it has swept up its own mass 
in the \ion{H}{2} region at a radius of $\sim$ 20 pc.  The 'broken sawtooth' velocity 
profile at 2.9 and 14.5 days soon relaxes into a new free expansion with a peak velocity 
of 55000 km s$^{-1}$.

\subsubsection{500 yr $< t <$ 6350 yr: End of the Free Expansion}

At 515.6 yr a homologous free expansion can still be seen in the density profile, which 
retains a flat central core and power-law dropoff.  At this point its radius is only 7.5 
pc and the blast has not yet accumulated its own mass in the relic \ion{H}{2} region.  By 
3093 yr it grows to 21 pc and has swept up 260 $\Ms$.  This marks the end of the free 
expansion:  the density profile flattens out and and a reverse shock forms at $r =$ 16 pc.  
The reverse shock strengthens, backstepping through and completely ionizing the remnant as 
shown in Figures \ref{fig:260Ms_halo1_1}f and \ref{fig:260Ms_halo1_1}h.  As this reverse 
shock propagates backward through the ejecta from 500 - 6000 yr in Figure
\ref{fig:260Ms_halo1_1}e, collisional ionization losses in H and He rise over approximately
the same interval of time in Figure \ref{fig:losses}g.  We note that at 6347 yr the density 
profile is reminiscent of the Chevalier phase in galactic SN remnants: a forward shock and 
reverse shock separated by a contact discontinuity \citep{chev74}. Inverse Compton scattering 
with the cosmic microwave background (CMB) begins to remove energy from the hot, ionized 
interior of the blast after 1 kyr as it grows in volume.  Compton losses reach 20\% of the 
energy of the explosion in this run, raising the possibility of small scale fluctuations 
in the CMB that might be observable through the Sunyaev-Zel'dovich (SZ) effect.  We examine 
this issue in greater detail in $\S$ 5.

In this era, the remnant is still well short of the dense shell, and has thus evolved in 
a uniform medium, as shown in the left panel of Figure \ref{HIIregions}, which shows the 
density profile of the \ion{H}{2} region.  Although not included in Figures 
\ref{fig:260Ms_halo1_1} and \ref{fig:260Ms_halo1_2} for brevity, reflected shocks traverse 
the interior of the remnant twice before it reaches the shell.  They are less important to 
the dynamics of the remnant than the adiabatic expansion against the surrounding material,
but as will be shown later, they never permit the flow to fully settle into an adiabatic
Sedov-Taylor solution.  By the time the remnant collides with the shell it has lost 70\% 
of its kinetic energy.  

\subsubsection{19.8 kyr $< t <$ 420 kyr: Collision with the Shell and the Radiative Phase}

The 400 km s$^{-1}$ shock overtakes the 25 km s$^{-1}$ \ion{H}{2} region shell at $r =$ 85 
pc at 59.5 kyr.  Its impact is so strong that a second reverse shock forms and separates 
from the forward shock at 420 kyr.  Both shocks are visible in the density and velocity 
profiles of Figure \ref{fig:260Ms_halo1_2}a at 175 and 210 pc. In reality, the interaction 
of the SN and shell is more gradual:  the remnant encounters the tail of the shell at 60 pc 
at 19.8 kyr, at which time the greatest radiative losses begin, tapering off by 7 Myr with 
the formation of another reverse shock.  Hydrogen Ly-$\alpha$ radiation dominates, followed 
by inverse Compton scattering, collisional excitation of He$^+$ and bremsstrahlung, but the 
remnant also collisionally ionizes H and He in the dense shell as evidenced by the sharp 
rise in ionization losses at 40 kyr.  The rise in total kinetic energy at late times is due 
to the appearance of the dense shell of the \ion{H}{2} region on the grid at 60 kyr, which 
cannot be separated from the remnant.  

\subsubsection{$t >$ 2 Myr: Dispersal of the Halo} \label{sec:shells}

Hydrodynamical profiles of the remnant at 2 Myr and at 7.93 Myr, the time to which the 
simulation was run, appear in Figures \ref{fig:260Ms_halo1_2}e-h.  Figure \ref{fig:losses}g
reveals that H Ly-$\alpha$ radiation drains the kinetic energy of the remnant out to 7 Myr 
and that collisional ionization continues to 4 Myr, well after the remnant has struck the 
shell.  These additional losses occur as the supernova overtakes a series of smaller 
subsidiary shocks in the relic \ion{H}{2} region visible just beyond the dense shell 
in the velocity profile of the left panel of Figure \ref{HIIregions}.  One feature of 
these secondary collisions is the formation of a dense shell at 500 pc at 2 Myr that 
persists to 800 pc.  The large recombination rates in this shell drop its ionization 
fraction to 0.08.  Since densities elsewhere in the remnant are nearly 100 times smaller, 
this shell is likely the site of the H Ly-$\alpha$ losses for $t >$ 2 Myr, which grow to 
50\% of the original energy of the blast.

The density plot at 7.93 Myr clearly shows that nearly all the baryons are ejected from 
the halo at velocities well in excess of the 2 - 3 km s$^{-1}$ escape speed.  The blast 
scours gas from inside the virial radius, to densities much lower than in the \ion{H}{2} 
region: 10$^{-4}$ cm$^{-3}$ as opposed to 0.1 cm$^{-3}$. Only 10\% of the energy of the 
blast survives radiative, gravitational, and adiabatic losses, but the momentum associated 
with this energy, together with strong relic ionized flows, still easily unbinds the gas 
from the halo.  In these circumstances the dispersal of heavy elements into the surrounding
intergalactic medium (IGM) is arrested when the remnant comes to pressure equilibrium with 
the relic ionized gas, typically at half the radius of the progenitor \ion{H}{2} region 
(one or two kpc) \citep{get07}. 

Both Compton and free-free losses from the remnant decline after 3 $\times$ 10$^{7}$ yr, 
in general agreement with their respective cooling timescales at these densities, 
temperatures, and redshift \citep{ky05}
\begin{equation}
t_{\mathrm{ff}} \sim 10^{7} \left(\displaystyle\frac{n_{\mathrm{H}}}{cm^{-3}}\right) \left(\displaystyle\frac{T_{\mathrm{e}}}{10^{7}K}\right) \mathrm{yr}, 
\label{eq:ff}
\end{equation} 
and
\begin{equation}
t_{\mathrm{IC}} \sim 7 \times 10^{6} \left(\displaystyle\frac{1+z}{20}\right)^{-4} \mathrm{yr}. \label{eq:IC}
\end{equation}
At 2 Myr, $n_{\mathrm{H}} \sim$ 0.1 cm$^{-3}$, $T_{\mathrm{e}} \sim$ 10$^5$ K, and $z \sim$ 20, so $t_{\mathrm{ff}}$ and 
$t_{\mathrm{IC}} \sim$ 10$^7$ yr.

The other six supernovae that explode in \ion{H}{2} regions follow nearly identical 
evolutionary paths.  Most of the initial blast energy goes into adiabatic expansion 
and then radiation upon impact with the ionized core shock in the fossil \ion{H}{2} 
region from 10$^4$ - 10$^5$ yr.  This timescale is set by the dynamics of the \ion{H}{2} 
region: the core shock radii fall within a fairly narrow band set by the sound speed of 
the ionized gas, which is a weak function of temperature and varies little for the three 
types of progenitor.  Radiative losses at these later times originate from both the 
remnant and the dense shell of the \ion{H}{2} region.  In all seven cases the SN destroys 
the halo: baryon densities interior to the virial radius fall below 10$^{-4}$ cm$^{-3}$.

\subsection{Blasts in Neutral Halos}  

\begin{figure*}
\epsscale{1.17}
\plottwo{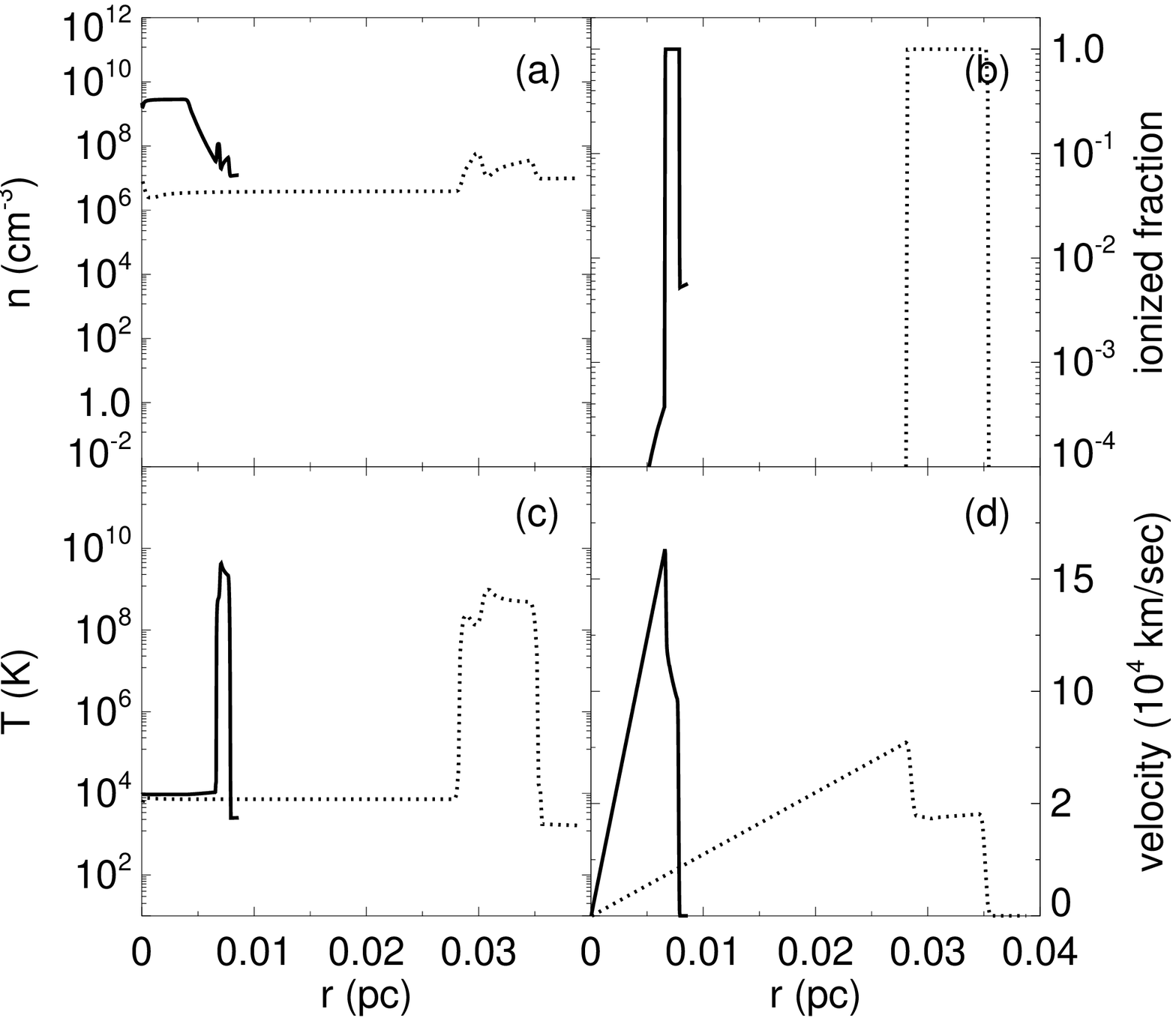}{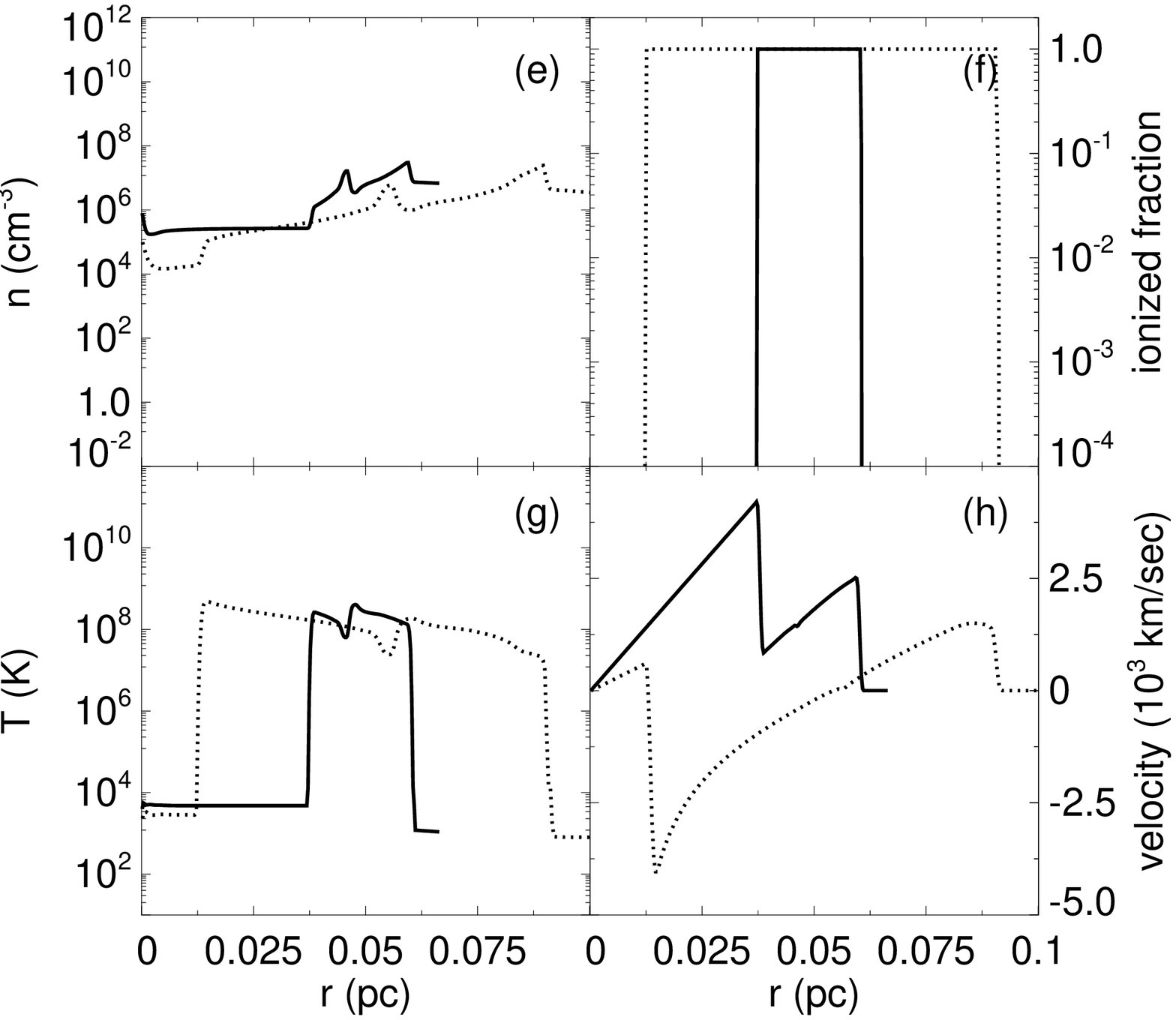}
\caption{Early evolution of the 40 $\Ms$ hypernova in halo 3.  Panels (a) 
- (d): early free expansion.  Solid: 0.301 yr; dashed: 3.01 yr.  Panels (e) 
- (h): formation of the first reverse shock. Solid: 7.45 yr; dashed: 17.4 yr.
\label{fig:40Ms_halo3_1}}  
\end{figure*}

\begin{figure*}
\epsscale{1.17}
\plottwo{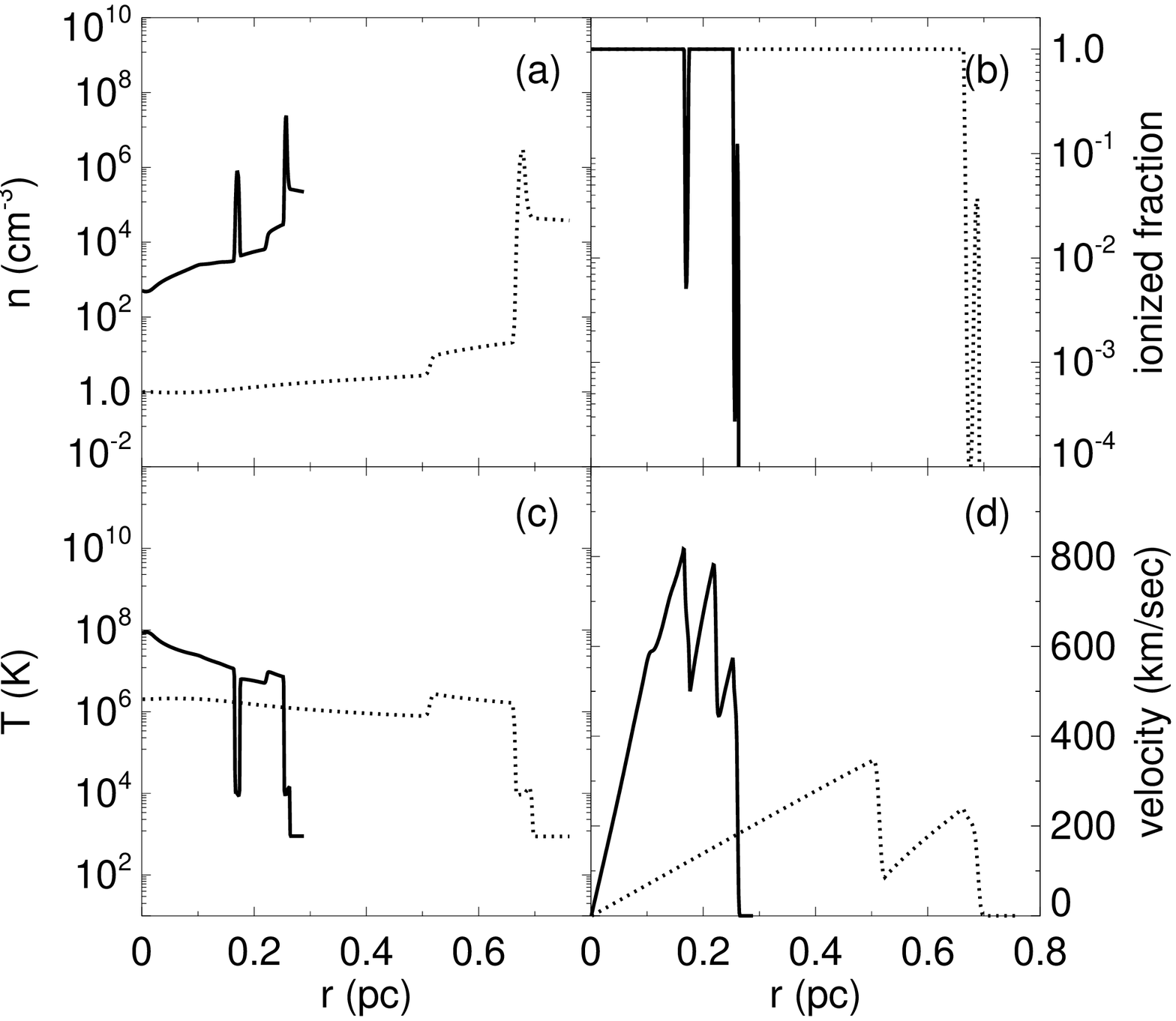}{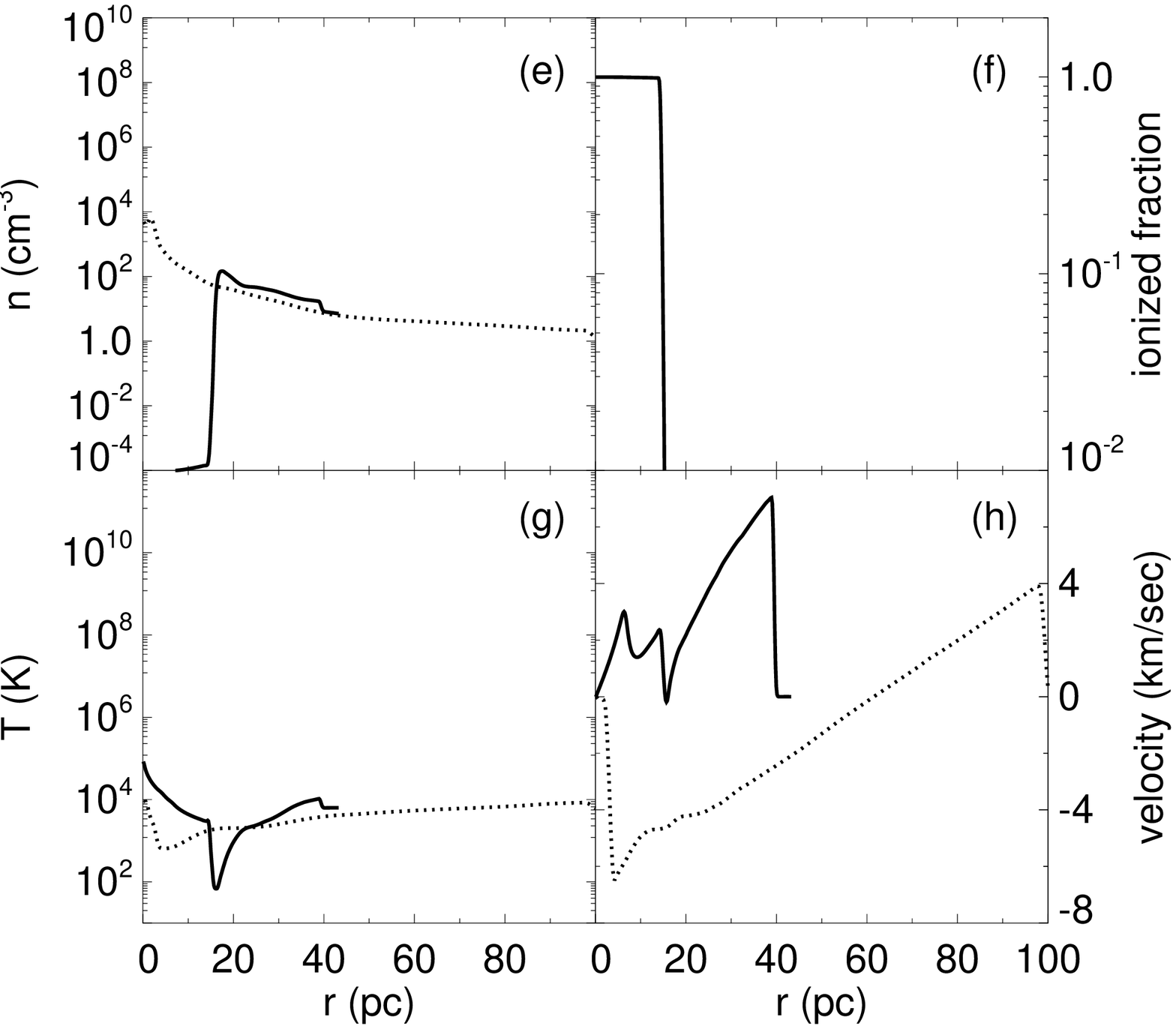}
\caption{Later evolution of the 40 $\Ms$ hypernova in halo 3. Panels (a) - 
(d):  multiple strongly radiatively-cooling shocks. Solid: 214 yr; dotted: 1348 yr.  
Panels (e) - (f): fallback of the remnant to the center of the halo.  Solid: 2.14
Myr; dotted: 6.82 Myr.
\label{fig:40Ms_halo3_2}}  
\end{figure*}

We show in Figures \ref{fig:40Ms_halo3_1} and \ref{fig:40Ms_halo3_2} four stages of evolution
for a 40 $\Ms$ hypernova in halo 3, the most massive halo. The \ion{H}{2} region is trapped 
so close to the star that the halo is essentially neutral and undisturbed when the star 
explodes.

\subsubsection{$t <$ 1 yr: Free-Expansion Driven Shock}

As before, the leading edge of the expansion promptly heats to extremely high temperatures,
this time to approximately 10$^{10}$ K.  This is somewhat less than in the PISN because of  
the inertia of the surrounding medium.  Although the edge of the shock is again completely 
ionized, it fails to jet forward as before because the ambient densities are so large, 
10$^{7}$ cm$^{-3}$ instead of 0.1 cm$^{-3}$.  The shock strongly radiates, primarily by 
bremsstrahlung and excitation of He, although the latter falls off rapidly as shown in Figure 
\ref{fig:losses}f.  Unlike PISN in the \ion{H}{2} region, this blast loses large amounts of 
energy at very early times, 3 $\times$ 10$^{50}$ erg by 1 yr.  At 0.301 yr the density profile 
is still mostly that of a homologous free expansion, but by 3.01 yr it has leveled out and a 
reverse shock can be seen in the velocity at 0.03 pc.  At this radius the blast has swept up 
several times its own mass in the neutral halo. The abrupt rise in collisional excitation and 
ionization losses in He$^+$ at $t \sim$ 1 yr coincides with the breakthrough of the dense shell, 
at 0.03 pc in Figure \ref{fig:40Ms_halo3_1}a, by the reverse shock, which requires 10$^{50}$ erg.

\subsubsection{$t <$ 20 yr: Early Radiative Phase}

The second jump in He$^+$ excitation and ionization cooling at 20 yr in Figure \ref{fig:losses}f
occurs when the reverse shock sweeps through and completely ionizes the dense interior of the 
fireball.  From $t =$ 2 - 20 yr, both adiabatic expansion and radiation drain two thirds of the 
energy from the blast in roughly equal proportions.  The remnant exhibits the first of several 
Chevalier phases at 7.45 yr: a forward and reverse shock separated by a contact discontinuity is 
visible in Figure \ref{fig:40Ms_halo3_1}e.  Bremmstrahlung losses level out after 20 yr because 
over the previous 10 yr the density of the shocked gas falls by a factor of 5 and cools below 
10$^{8}$ K.  This is again consistent with the free-free cooling time predicted by eq \ref{eq:ff} 
for $n_{\mathrm{H}} \sim$ 2 $\times$ 10$^6$ cm$^{-3}$ and $T_{\mathrm{e}} \sim $ 3 $\times$ 10$^9$ 
K at 0.3 yr.

\subsubsection{100 $< t <$ 5000 yr: Late Radiative Phase}

The inertia of the neutral halo rapidly slows the blast, allowing time for a series of 
reflecting shocks to reverberate between the edge of the blast and the center of the 
interior.  Reverse shocks continue to break off from the leading shock as it plows up 
more of the neutral halo and is overtaken by reflected shocks emerging from the interior. 
Flow profiles of this process are shown in Figures \ref{fig:40Ms_halo3_2}a-d.  At 214 yr 
a new reverse shock in the density profile at 0.25 pc is about to break away from the 
forward shock as a reflected shock approaches it from behind.  At 1350 yr these two 
structures merge into a new single Chevalier phase with another reverse shock about to 
depart toward the origin.

These crossing shocks form dense shells that rapidly recombine and become mostly neutral.  
The shells are luminous sources of H Ly-$\alpha$ emission, which cools them to 10000 K 
amidst the otherwise fully ionized and very hot gas.  They are the sites of the H Ly 
$\alpha$ losses from the remnant from 50 to 5000 yr in Figure \ref{fig:losses}f that eventually 
amount to a third of the energy of the blast.  In this period the kinetic energy of the 
remnant falls to 0.5\% of the energy of the SN but at 5000 yr and 1.5 pc it still has 
sufficient momentum to expand.

\subsubsection{Fallback}

We show in Figures \ref{fig:40Ms_halo3_2}e-h the remnant at 2 Myr and 7.93 Myr.  At 2 Myr
the remnant has devolved into a slow dense shell of warm neutral gas extending from 15 
to 40 pc and surrounding a hot, ionized diffuse ($\lesssim$ 10$^{-4}$ cm$^{-3}$) interior. 
The inner surface of the shell soon falls back into the dark matter potential as shown at 
7.93 Myr in \ref{fig:40Ms_halo3_2}e.  The outer region of the shell ripples outward but at 
velocities well below that required for escape from the halo.  By now the gas is fully 
neutral.

In contrast to blast models that 'fizzle' when initialized by thermal energy in massive
neutral halos \citep{ky05}, the free expansion in our model delivers enough impulse to 
the halo to seriously disrupt it out to 40 pc, even though the remnant retains less than 
1\% of the energy of the blast by 2 Myr.  Although the baryons remain bound to the halo, 
the ejecta heavily enriches them with metals prior to fallback, radically altering their 
cooling and fragmentation timescales.  We find that the baryons can rebound from the center 
more than once after their initial recollapse into the halo, with large, episodic infall 
rates onto the central black hole.  This may constitute a mechanism for the rapid growth 
of supermassive black hole seeds at high redshift.  The explosion and its metals remain 
completely confined to the halo, unable to enrich the early IGM.

As another example of fallback in a trapped explosion, the evolution of the 15 $\Ms$ 
supernova in halo 3 is qualitatively similar to that of the hypernovae, but with a few 
differences.  First, the free expansion slows more quickly because it has less energy.  
The reverse shock at first fails to fully detach from the forward shock and instead 
forms a dense shell that emits extremely intense He lines.  By approximately 1.1 yr 
this radiation slows the forward shock, cooling it suddenly from 1 $\times$ 10$^{8}$ to 
10$^{4}$ K in just 3.5 yr, extinguishing bremsstrahlung x-rays at 6 yr and terminating 
all other radiative losses by $t \sim$ 20 yr.  Afterwards, the blast loses kinetic energy 
only to adiabatic expansion and gravity.  The remnant passes through a single Chevalier 
phase at $t \sim$ 6000 yr, after which the reverse shock propagates back to the center 
of the halo.  The reverse shock is too weak to ionize the interior, and the reverberations 
back to the forward shock are too faint to form shells.  The SN therefore remains mostly 
neutral.  Gravity arrests its expansion at 2.5 pc at 1.35 $\times$ 10$^{5}$ yr, and it 
falls back to the center of the halo.  Helium lines from the semi-detached reverse shock 
shell dominate the radiation from this remnant, emitting 75\% of the energy of the blast.

\begin{figure}
\resizebox{3.45in}{!}{\includegraphics{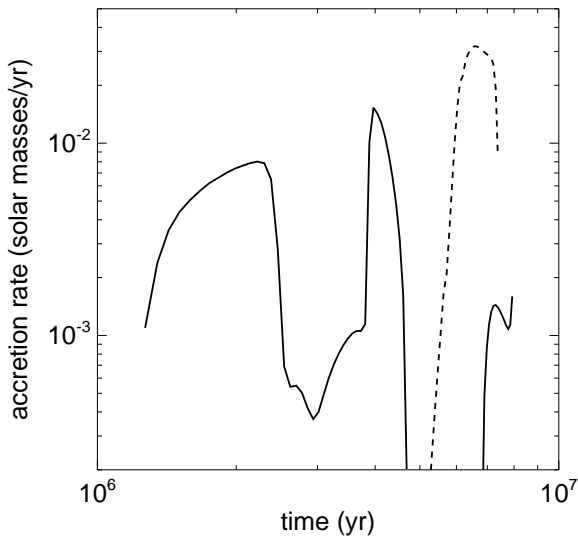}}
\caption{Infall rates $4 \pi \rho(r) v(r) r^2$ associated with fallback of the 15 
and 40 $\Ms$ remnants in halo 3. Solid: 15 $\Ms$; dashed: 40 $\Ms$} 
\label{fig:accr}
\end{figure}

We show central infall rates associated with the fallback of the 15 and 40 $\Ms$ 
remnants in halo 3 from 10$^6$ yr to 2 $\times$ 10$^7$ yr in Figure \ref{fig:accr}.  Infall
occurs later in the more massive remnant because its greater explosion energy carries it
farther from the center of the halo.  There are three fallback phases in the 15 $\Ms$
remnant over this time interval ranging from 10$^{-3}$ - 10$^{-2}$ $\Ms$ yr$^{-1}$ 
in magnitude and from 1 - 2 Myr in duration.  The three episodes are punctuated by 
rebound from the core of the halo, which is not shown in the figure.  The first infall
phase of the 40 $\Ms$ remnant manifests rates of 3.5 $\times$ 10$^{-2}$ 
$\Ms$ yr$^{-1}$ and lasts 1.5 Myr.  These are not the only occurences of fallback;
our numerical models indicate several more will occur in less than a merger time.  In 
addition to fueling rapid growth of a black hole, these episodes will thoroughly mix 
metals from the ejecta in the central regions of high density in the halo because the 
leading shocks are subject to dynamical instabilities during both rebound and collapse.
However, we point out that fallback of the remnant to the center of the halo is distinct
from accretion onto the black hole, whose rates are governed by angular momentum and 
radiation transport away from the central object.  Future radiation hydrodynamical 
simulations will determine the rates which the black hole grows in mass.

Intense bursts of x-rays from the central black hole will likely accompany these periods
of heavy accretion, ionizing and heating gas within the halo and perhaps driving strong
H$_2$ formation.  It is generally held that the broad I-fronts of black holes and quasars 
exert positive feedback on structure formation because H$_2$ formation is promoted in their 
partially ionized outer layers \citep[e.~g. ][]{rs01}.  However, this picture is complicated
by the fact that the accretion shock might radiate strongly in UV, including the LW bands,
which would suppress H$_2$ formation.  Furthermore, recent work on the structure of quasar
ionization fronts indicates that they are preceded by a 'T-front', or temperature front 
\citep{qiu08}.  This is a layer of 10$^4$ K minimally-ionized gas formed beyond the broadened 
I-front of black holes and quasars by the few most energetic photons.  These photons cause 
secondary ionizations in the pre-front gas that heat it but cannot sustain a large ionization 
fraction.  The preheated gas might pre-empt H$_2$ formation that would otherwise occur when 
the outer layers of the I-front reaches it.  Detailed numerical simulations will be required
to resolve the thermal and chemical evolution of the gas in the vicinity of the seed black 
holes at the centers of large halos at high redshifts.

The importance of including the dark matter gravitational potential in primordial SN models 
is particularly clear in trapped explosions.  We performed experiments in which we set the 
halo in hydrostatic equilibrium and then evolved the blast without force updates by the dark 
matter potential.  In these tests, the remnant simply continues expanding and there is no 
fallback.  This provides a good check of the gravitational physics on the expanding grid 
and attests to the necessity of including dark matter in the calculation.

\subsection{Halo Destruction Efficiency}

\begin{figure*}
\epsscale{1.17}
\plottwo{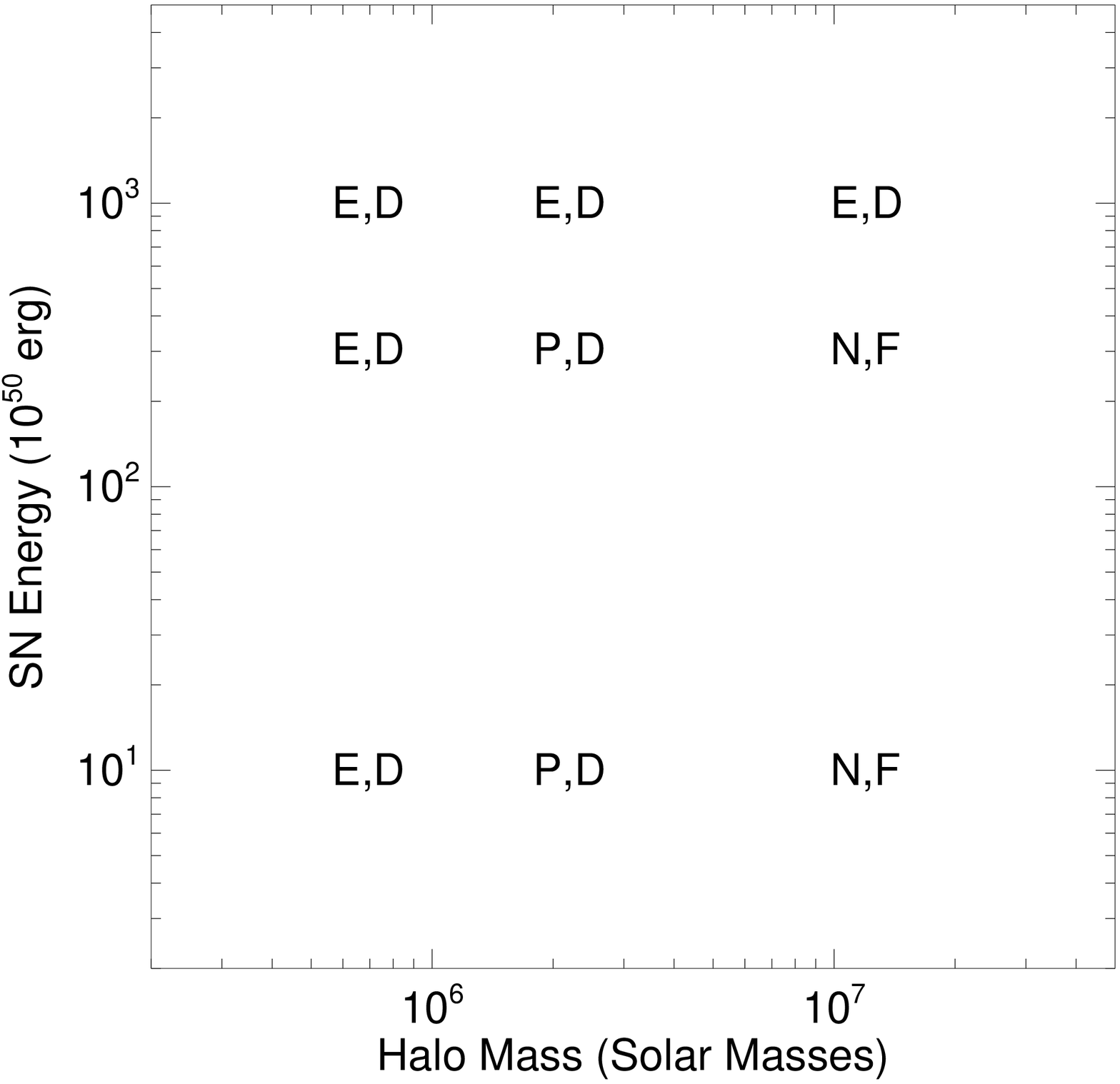}{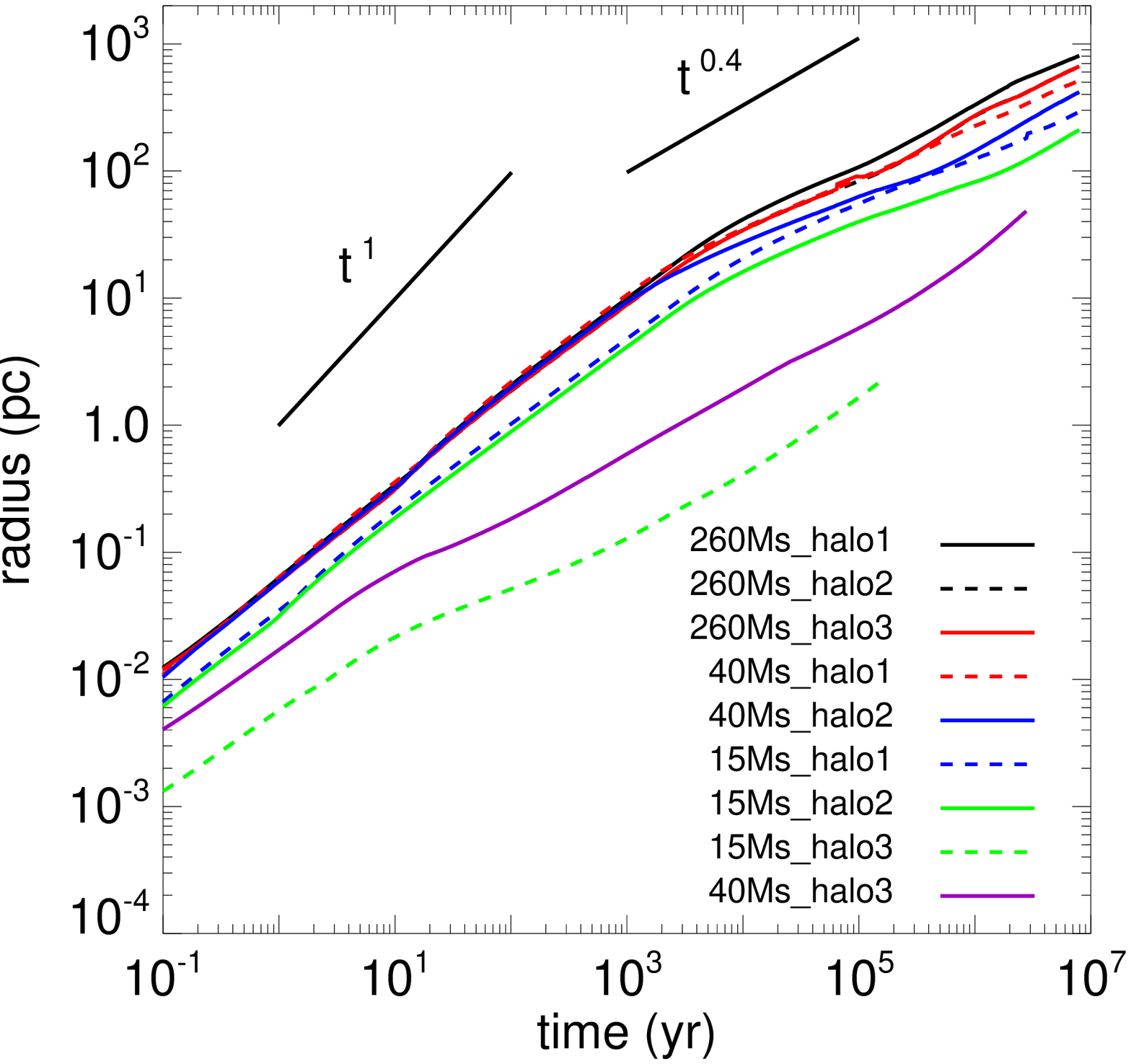}
\caption{Left: eventual fate of a halo given the indicated explosion energy.  The first
letter refers to the final state of the halo prior to the explosion; E: photoevaporated;
P: partly ionized, defined as the I-front not reaching the virial radius; N: neutral, or
a failed \ion{H}{2} region.  The second letter indicates outcome of the explosion; D: 
destroyed, or F: fallback.  Right: shock position $vs$ time for the nine supernova remnants.
\label{fig:RS}}  
\end{figure*}

We summarize the results of our explosion models in the left panel of Figure \ref{fig:RS}.   
Halos less massive than 2.1 $\times$ 10$^6$ $\Ms$ are destroyed by Population III stars as 
small as 15 $\Ms$ (we reiterate that destruction of the halo refers to expulsion of its gas, 
not its dark matter).  On the other hand, PISN disperse much larger halos, even the first 
that could cool by atomic lines ($\sim$ 1.0 $\times$ 10$^7$ $\Ms$).  This in part is because 
the star preionizes the halo.  Indeed, even lower-mass primordial stars finally eject more 
than 90\% of the baryons from halos $\lesssim$ 10$^7$ $\Ms$ because their \ion{H}{2} regions 
impart considerable momentum to the baryons prior to the explosion that facilitates breakout 
of the remnant into the IGM.

In failed \ion{H}{2} regions, enough momentum from even modest explosions, 10$^{51}$ erg and 
3 $\times$ 10$^{52}$ erg, survives the enormous prompt radiative losses to disrupt halos out 
to a significant fraction of their virial radii.  Nevertheless, we find that Type II supernovae 
and hypernovae are contained by halos more massive than 1.0 $\times$ 10$^7$ $\Ms$, but will 
destroy less massive halos that cool mainly by molecular hydrogen.  Recall that many primordial 
stars never explode: those lying between 50 and 100 $\Ms$ and above 260 $\Ms$ are not thought to 
exhibit any kind catastrophic mass loss \citep{hw02}.  In these circumstances the halos are 
instead photoevaporated, with the loss of $\sim$ 50\% of the gas rather than 95\%, as expected 
for a SN.  Although not all gas exits these halos, no subsequent star formation would follow for 
at least a merger time.  

\subsection{Comparison to the Canonical Phases of Idealized Remnants}

In the classic problem of a supernova blast in a uniform medium, idealized remants evolve
in four distinct phases: the free expansion, the Sedov-Taylor (ST) phase, the pressure 
driven snowplow (PDS), and the momentum conserving snowplow (MCS).  Each stage can be 
loosely associated with a power law for the position of the shock as a function of time:  
$R_{\mathrm{S}} \propto t^{\eta}$.  In the early free expansion the flow is completely dominated by 
the momentum of each fluid element and this exponent is 1.  Later, after the blast has 
swept up many times its own mass it consists of a self-similar adiabatic hot interior and 
non-radiative shock: the ST phase \citep{sed59,tay50} in which $\eta_{\mathrm{ST}} = 2/5$.  As the 
remnant grows, the shock begins to be driven by the pressure of its hot, now nearly isobaric 
interior in addition to its own momentum because the postshock fluid velocity approaches 
that of the shock \citep{cox72,chev74}.  At about the same time, if the shocked gas can cool 
by radiation it collapses into a thin dense shell that accrues material as it 
'snowplows' the interstellar medium (ISM).  If the interior cannot radiatively cool in this 
pressure-driven snowplow (PDS) era its shock evolves according to $\eta_{\mathrm{PDS}} = 2/7$ \citep{mo77}.  
Finally, if the interior loses energy both to $PdV$ work and radiation, the remnant becomes 
a momentum conserving snowplow (MCS) whose analytical solution yields $R_{\mathrm{S}} \propto t^{1/4}$ 
\citep{oort51}.  The shell halts when it comes into pressure equilibrium with the ISM, after
which it disperses and becomes indistinguishable from its surroundings.  

We show in the right panel of Figure \ref{fig:RS} shock position as a function of time for 
all nine remnants.  All clearly exhibit an $\eta \sim 1$ expansion at early times but depart 
from the canonical stages of ideal remnants in uniform media thereafter.  This is not surprising 
for several reasons.  When the reverse shock breaks away from the contact discontinuity at the
end of the free expansion, it reverberates back and forth across the interior, heating it and 
never allowing it to relax to an ST phase, even though the leading shock does not radiate. This
is likely why $\eta < $ 2/5 after the free expansion.  When the remnant later mingles with the 
baryons that have been swept up into the \ion{H}{2} region shock, it becomes strongly radiating 
without the formation of a thin shell.  The curves taper somewhat more as the shock deaccelerates 
upon encountering the shell, as seen from 10$^{4}$ to 10$^{5}$ yr.  The remnant cannot be 
disentangled from the strong fossil ionized flows thereafter.  

On the other hand, supernovae in the dense cores of neutral halos depart from the standard
behavior of ideal remnants in uniform densities because the cores are so stratified (with
baryon densities falling by $r^{-2.2}$), and because they radiate strongly from very early
times.  As a result, the free expansion ends in less than a year, and shocks reverberate 
through interior of the remnant even more frequently than in \ion{H}{2} regions due to the 
inertia of the surrounding gas, again preventing the ST phase from materializing.  Two stages 
of radiative loss without the formation of thin shells in the metal-free shocked gas then 
follow. Through all three stages adiabatic expansion also slows the remnant, with eventual 
fallback into the halo. Note that the plots in Figure \ref{fig:RS} trace the position of the 
leading velocity front in the gas, which can continue outward into the halo even if the bulk 
of the remnant, which does not have a precise location, falls back toward the center. In short, 
we find that primordial supernovae do not conform to the usual paradigm for ideal remnants in 
uniform media.

Note from these profiles that the shock experiences brief periods of acceleration in neutral
halos and may therefore be susceptible to Rayleigh-Taylor instabilities in three dimensions.  
Furthermore, the contact discontinuities separating forward and reverse shocks would also be 
prone to instabilities and breakup in multidimensional simulations \citep{mo77}.  

\section{Observational Signatures of Primordial Supernovae}

One trend that is immediately apparent in all nine remnants in Figure \ref{fig:losses} is 
that adiabatic expansion losses precede radiative losses in blasts evolving in \ion{H}{2} 
regions but are coincident in neutral halos.  Another is that bremsstrahlung dominates 
energy losses in explosions in neutral halos but amounts to at most a few percent in 
\ion{H}{2} regions.  Inverse Compton scattering is greatest in the most powerful explosions 
in \ion{H}{2} regions but virtually nonexistent in neutral halos, where the fireball cools 
before it can enclose large volumes of the CMB. In \ion{H}{2} regions, H Ly-$\alpha$ finally 
dominates radiative losses from primordial remnants, with singly-ionized He excitation next 
in importance.  Neutral He emission rivals or exceeds H Ly-$\alpha$ in less energetic blasts 
in neutral halos because He is not completely ionized.  We plot luminosities due to excitation 
of H and He along with bremsstrahlung and IC for all nine remnants in Figure \ref{fig:luminosities}.

\subsection{Luminosities in \ion{H}{2} Regions}

\begin{figure*}
\epsscale{1.15}
\plotone{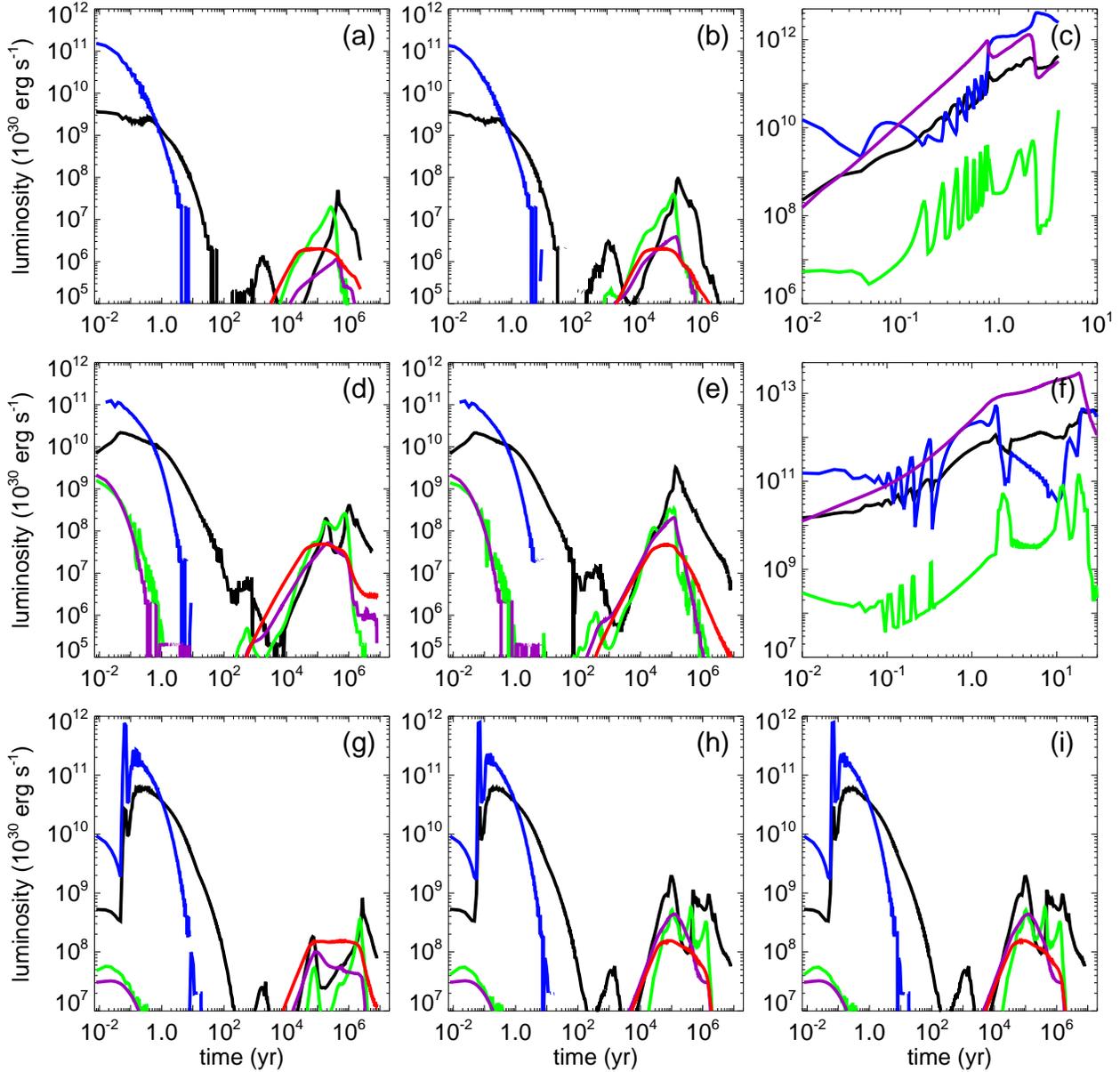}
\caption{Luminosities of all nine primordial SN remnants $vs$ time.  Black: collisional 
excitation of H; blue: collisional excitation of He; green: collisional excitation of 
He$^+$; purple: x-rays (bremsstrahlung); red: energy deposition rates into the CMB due
to inverse Compton scattering.  Panels (a) - (c):  15 solar mass Type II SN in halo 1, 
halo 2, and halo 3, respectively. Panels (d) - (f):  40 solar mass hypernova in halo 1, 
halo 2, and halo 3, respectively.  Panels (g) - (h):  260 solar mass pair-instability 
SN in halo 1, halo 2, and halo 3, respectively.  Note the different energy and time
scales in panels (c) and (f), which describe the trapped explosions.  As discussed in
the text, the H and He line luminosities in the first pulse in H II region explosions
should really be considered free-free losses.
\label{fig:luminosities}}  
\end{figure*}

Type II supernovae and hypernovae in \ion{H}{2} regions emit three flashes of Ly-$\alpha$ 
lines: the prompt flash when the free expansion forms a shock with its surroundings, a 
second flash as the reverse shock breaks free of the forward shock and ionizes the interior, 
and a final flash when the remnant collides with the dense \ion{H}{2} region shell.  Peak 
luminosities in the second and third flashes scale approximately with explosion yield: 10$^
{38}$ erg s$^{-1}$ for the Type II supernova and 10$^{39}$ erg s$^{-1}$ for the hypernova 
and PISN.  The third flash in more powerful explosions has twin peaks due to collisions with 
the smaller subsidiary shocks that lie just beyond the dense shell in the \ion{H}{2} regions 
of more massive progenitors, as discussed in section \ref{sec:shells}.  The contributions to 
each flash by H and He$^+$ are in proportion to their abundances.  Because He is always singly
ionized at later times it is absent in the second and third flashes.  The second flash is 
consistently two orders of magnitude less intense than the third flash because the mass of the 
remnant is much lower than that of the dense shell and remnant combined.

It has been suggested that Ly-$\alpha$ photons from Population III stars are scattered by H 
in the early IGM, coupling its spin temperature to its kinetic temperature without heating 
the gas \citep{cm04}.  If the scattering region is not first heated by hard photons, the spin 
temperature would rapidly fall to the low kinetic temperature of the gas and the region would 
appear in absorption in 21 cm against the CMB.  If more energetic photons later heat the 
scattering volume it would then appear in emission against the CMB, creating a brief 
absorption signal followed by a weak emission signal that is unique to Population III stars. 
Our findings indicate that the same may be true of the supernova remnant but in reverse:
x-rays from the collision of the remnant and the dense shell die out before the Ly-$\alpha$
pulse, resulting in a strong 21 cm absorption signal preceded by weak emission against the 
CMB.  Given that the lifetime of the main flash can exceed that of the star and that the
scattering region can span several kpc proper, it may be possible to resolve them with the 
next generation of 21 cm observatories such as the \textit{Square Kilometer Array} 
(\textit{SKA}).

Although the Ly-$\alpha$ intensity of the first flash is at least a hundred times greater 
than in the next two, the actual energy released is minute due to its transience and is of 
little importance to the evolution of the free expansion.  The large fluxes are in part 
numerical; they are emitted from the extremely large densities at the interface of the 
shock and the free expansion, which are not spatially resolved by the mesh.  Artificial 
viscosity smears the enormous jump in density there over several zones, over which the 
temperature rises from a few thousand to several billion K.  In one of the zones the 
temperature crosses the threshold for H and He line emission, activating these cooling 
channels for the entire zone when in reality they originate from only an extremely thin
shock layer that is a tiny fraction of the width of the zone.  The Ly-$\alpha$ luminosity 
is thus overestimated by a factor roughly equal to the ratio of the widths of the zone and 
the shock; it is probably several orders of magnitude lower than free-free emission for
$t < $ 10 yr.  Little inaccuracy is introduced to the energetics of the ejecta by this 
artifact, not only because the energies involved are small.  If excitational cooling 
of H and He are deactivated in this phase the shock simply remains hotter, with elevated
free-free emission rates that are nearly identical to the total rates when excitational 
cooling is included.  Thus, nearly all the radiation in this era is really in x-rays, with 
peak luminosities of 10$^{41}$ erg s$^{-1}$.  Because the energy liberated is less than a 
millionth of the blast energy it does not appear in Figure \ref{fig:losses}.  All three types 
of supernovae have similar luminosity profiles in \ion{H}{2} regions in this era because the 
ejecta was assumed to have the same initial velocity profiles.  Note that the energy of 
the first pulse, which is from the interaction of the shock and ambient \ion{H}{2} region, does 
not include the energy of radioactive decay in the ejecta, which is expected to be similar 
to that observed in SN today, $\sim$ 10$^{42}$ erg s$^{-1}$.

Hypernovae and PISN in \ion{H}{2} regions emit a second x-ray flash during the collision with 
the dense shell.  In contrast to the prompt flash, the second burst is longer, $\sim$ 
10$^6$ yr with luminosities of 10$^{38}$ erg s$^{-1}$.  Type II 
supernovae emit only the second flash, with emission rates that are lower by two orders 
of magnitude.  The durations of the second x-ray pulse and the inverse Comptonization of 
the CMB are again in agreement with the free-free and IC cooling times of eqs \ref{eq:ff} 
and \ref{eq:IC} for densities of 0.1 cm$^{-3}$ and temperatures of 10$^7$ K at 1 Myr.  
Although primordial supernovae were the first x-ray sources in the early universe they 
probably did not contribute to cosmological reionization \citep{tse93}.  After examining 
early reionization by x-rays released by primeval supernovae, \citet{ky05} concluded that 
unrealistically large explosion rates are required to create significant x-ray backgrounds 
at high redshift.  However, prompt x-rays from blasts completely contained within neutral 
halos, although not visible to external observers, may drive strong H$_2$ formation that 
enhances the cooling and collapse of gas in the halos.

As noted earlier, Figures \ref{fig:losses} and \ref{fig:luminosities} show that cooling via 
inverse Compton scattering of cosmic microwave background photons contributes strongly to 
the loss of kinetic energy in supernova remnants evolving in \ion{H}{2} regions.  In some cases, 
such as the one shown in Figure \ref{fig:losses}g, it is more than 10\% of the overall energy 
loss, but more generally only a few percent of the supernova's kinetic energy is deposited 
into the CMB.  The strength of this cooling component is due to the high energy density of 
the CMB at the redshifts of these SN, and has been predicted by previous work 
\citep{tse93,v96,met01}.  One result will be a distortion of the CMB due to the 
Sunyaev-Zel`dovich effect, predicted to be on the order of $\delta y \sim \mathrm{few} \times 10^{-6}$  
\citep{ock03}.  The angular size of a given HII region and supernova remnant at $z \sim 20$ 
is quite small, however -- on the order of 100 kpc (comoving), which corresponds to an 
angular size of a few arc seconds.  While individual Pop III supernovae would be essentially 
impossible to detect in the SZ with current telescopes, clusters of supernovae would create 
spectra distortions with a power spectrum that can be predicted by making assumptions about 
clustering bias.  We have examined the assumptions made by \citet{ock03}, and our results 
for radiative losses due to inverse Compton scattering are generally much lower than their 
predictions of 30-100\% of the supernova kinetic energy, suggesting that their results would 
be an upper bound on the contribution of Population III supernovae to the SZ effect.  Indeed, 
recent work by \citet{oshea07a} suggests that the vast majority of Population III stars will  
form in halos with virial temperatures below $10^4$~K, with a contribution to the angular 
power spectrum somewhere between cases B and C of Oh et al. (shown in Figure 1 of their paper).  
This implies that contributions to the SZ effect from Population III SN would be negligible in 
comparison to that from galaxy clusters, making them extremely difficult to detect.

\subsection{Radiation from Neutral Halos}

\begin{deluxetable}{cccc}
\tabletypesize{\scriptsize}
\tablecaption{Mass of Enriched Baryons\label{tbl-6}}
\tablehead{
\colhead{halo} & \colhead{Type II SN} & \colhead{Hypernova} & \colhead{PISN}}
\startdata
 1 & 8.59E+5 $\Ms$ & 1.75E+6 $\Ms$ & 5.35E+6 $\Ms$ \\
 2 & 4.35E+5 $\Ms$ & 1.61E+6 $\Ms$ & 3.38E+6 $\Ms$ \\
 3 & 8.84E+4 $\Ms$ & 1.61E+5 $\Ms$ & 3.38E+6 $\Ms$ \\
\enddata
\end{deluxetable}

Luminosity profiles for blasts in neutral halos exhibit a single peak lasting from 2 - 20 yr 
that is dominated by free-free and He line emission.  The duration and peak intensity both 
scale with explosion energy: 2 yr and 4 $\times$ 10$^{42}$ erg s$^{-1}$ for the Type II SN 
and 20 yr and 2 $\times$ 10$^{43}$ erg s$^{-1}$ for the hypernova.  Approximately 75\% 
percent of the blast energy is lost in this one burst of radiation, which emanates from a 
volume less than 0.1 pc in radius.  Again, these luminosities originate from the shock and
not the radioactive decay of the ejecta.  The luminosity of the shock in the Type II SN is
comparable to that due to decay in sn today.

Line emission from the 15 $\Ms$ and 40 $\Ms$ SNR in halo 3 exhibits rapid 
oscillations in the first year on periods of a few days.  This interesting phenomenon is
due to the reverse shock that is attempting to break free of the contact discontinuity.
Just as it does so, its densities and temperatures rise, with a sharp increase in line 
emission which then cools the shock.  Losing pressure support, the reverse shock 
retreats toward the contact discontinuity in the frame of the flow until gas from the 
interior again builds at the interface and causes the shock to break free, beginning the
cycle once more.  Oscillations in standing radiative shocks are well understood in a 
variety of astrophysical contexts \citep{ci82,iet84,anet97}.  The amplitudes of these variations
in our luminosity profiles clearly demonstrate the reverse shock to be the origin of 
almost all the atomic lines for $t <$ 1 yr.  Although not shown for brevity, oscillations 
are also present in the ionization cooling rates, again revealing the reverse shock to be 
their origin.  Corresponding variations in the x-rays are absent because they originate  
from the shocked gas on the other side of the contact discontinuity.  

Because the fireball cools fairly deep inside the halo, few of its x-rays are likely 
to escape into the IGM.  In ambient densities of 10$^7$ cm$^{-3}$, the mean free paths 
of 100 eV and 1 keV photons are 1.7 $\times$ 10$^{-7}$ and 3 $\times$ 10$^{-4}$ pc, 
respectively, so they would downscatter in energy many times before exiting the halo.
It is possible that the x-ray photons would promote the formation of molecular hydrogen  
in the suppressed remnant, but this would be transient and only operate at intermediate 
radii.  The resultant cooling would be much less important than the metal-line cooling 
near the center of the halo.  As shown in Figure \ref{fig:luminosities}, confined 
explosions deposit little energy into the CMB because they cool before enclosing an 
appreciable volume of background photons.  The visibility of atomic lines from deep 
inside the halo is also open to debate since the densities in the explosion envelope 
would cause many resonant scatterings for each photon.  The photons may downscatter 
to radio wavelengths, as observed in hypercompact \ion{H}{2} regions deep within 
molecular cloud cores \citep{church02,rod05}.  Detailed calculations of Ly-$\alpha$ 
radiative transfer are necessary to evaluate the 21 cm footprint of these explosions.

\subsection{Optical/IR Light Curves}

\begin{deluxetable}{cccc}
\tabletypesize{\scriptsize}
\tablecaption{Average Metallicity of Enriched Baryons\label{tbl-7}}
\tablehead{
\colhead{Halo} & \colhead{Type II SN} & \colhead{Hypernova} & \colhead{PISN}}
\startdata
 1 & 6.87E-3 $\Zs$ & 1.16E-2 $\Zs$ & 2.32E-2 $\Zs$ \\
 2 & 1.36E-2 $\Zs$ & 1.26E-2 $\Zs$ & 3.68E-2 $\Zs$ \\
 3 & 6.68E-2 $\Zs$ & 1.26E-3 $\Zs$ & 3.68E-2 $\Zs$ \\
\enddata
\end{deluxetable}

In our discussion we have neglected the optical/IR flash due to the radioactive decay of 
the ejecta, which would compete with the early radiation flash from the shock in both H
II regions and neutral halos.  Since nucleosynthetic yields are not well determined for 
these primordial events, we compare the peak optical luminosities of supernovae in the 
local universe to the first x-ray peaks in our simulations.  First, it is worthwhile to
note that at early times the ejecta is optically thick to its own decay photons, which
diffuse out through the envelope.  If we adopt the simple argument that the optical/IR
flash reaches its peak when the radiation diffusion timescale through the envelope is
equal to the age of the remnant, we find that \citep{a82,pe01}
\begin{equation}
t_{\mathrm{peak}} = \left(\frac{\kappa M_{\mathrm{ej}}}{7cv_{\mathrm{ej}}}\right)^{\frac{1}{2}},
\end{equation}
where M$_{\mathrm{ej}}$ and v$_{\mathrm{ej}}$ are the ejecta mass and peak velocity, 
respectively.  Commonly used values of the opacity $\kappa$ range from 0.1 - 0.4, 
depending on the inclusion of heavier element lines in addition to Thomson scattering.  
Assuming ejecta masses of 13 - 50 $\Ms$ and shock speeds of 25000 km s$^{-1}$, the light 
curves peak as early as 25 days and as late as 100 days, so they will compete with the 
early bremsstrahlung pulse.  

Type Ia supernovae peak luminosities are clustered around 9.6 $\times$ 10$^9$ $\Ls$
\citep[e.~g. ][]{bt92} but those of Type II supernovae vary from 0.4 - 4 $\times$ 
10$^9$ $\Ls$ \citep{ts90}.  Maximum x-ray luminosities in our models fall between
2.6 - 7.8 $\times$ 10$^7$ $\Ls$, hundreds of times lower than the optical/IR curves.
However, the x-ray pulse lasts for 2 - 4 yr, approximately ten times longer than the
decay flash.  We conclude that visible and IR radiation outshine the x-rays but that 
the x-rays outlive the optical glow.  As discussed in the previous subsection, atomic 
cascades in neutral halos likely downscatter these photons to much lower energies. 

\section{Chemical Enrichment}

We find the early emergence of a Chevalier phase in all our models to be a predictor 
of efficient metal mixing at small radii in both \ion{H}{2} regions and in neutral 
halos. In three dimensions, the susceptibility of the contact discontinuity separating 
the reverse and forward shocks to breakup \citep{ciof88} suggests that mixing would be 
prompt in neutral halos and be well underway by 25 pc in \ion{H}{2} regions.  However, 
the forward and reverse shocks may also exhibit Rayleigh-Taylor instabilities.  This 
would enhance mixing in the outer remnant and perhaps allow ambient gas to reach the 
center as the reverse shock propagates through the interior.  This is in contrast to
earlier work \citep{get07}, whose numerical resolution could not address these 
phenomena.  

How much gas is enriched by metals in these models, and to what levels?  In \ion{H}{2} 
regions that completely break out of the halo, the processes just described would 
enrich only half of the baryons originally interior to the virial radius of the halo, 
since the remainder are swept up into a dense shell 100 pc from the star.  The degree 
to which the shell would be contaminated is yet to be determined, and is the focus of 
three-dimensional multiscale calculations now under development.  Even more gas will 
be enriched as the shell and remnant expand together and then stall, typically at half 
the final radius of the \ion{H}{2} region.  In cases where the \ion{H}{2} region exits 
the halo, we estimate the mass and metallicity of the enriched gas by assuming the SN 
ejecta is uniformly mixed with all the baryons interior to the stall radius.  If the 
I-front ionizes only part of the halo, we approximate the final radius of the remnant 
from its deceleration beyond the virial radius and assume the ejecta to mix with all 
the gas inside this region.  Finally, in neutral halos disrupted by the blast we 
consider the gas displaced by the remnant prior to fallback to be well mixed, which 
will be true after repeated episodes of infall and rebound in the dark matter potential.  
Estimates of the mass and metallicity of enriched gas in each model are tabulated in 
Tables \ref{tbl-6} and \ref{tbl-7}.

Explosions in both \ion{H}{2} regions and neutral halos enrich 10$^{5}$ to 5 $\times$ 
10$^6$ $\Ms$ of gas with metallicities ranging from 7 $\times$ 10$^{-3}$ to 7 $\times$ 
10$^{-2}$ $\Zs$.  This is well above the critical metallicity believed to result in a 
rollover from high-mass to low-mass star formation in the early universe \citep{bcl01,
mbh03,ss07}.  There are two modes of chemical enrichment in our models.  Metals are 
either propagated far into the IGM, not returning to the halo in less than a merger 
time, which is $\sim$ 20 Myr at this epoch, or are confined deep within the parent halo.  
As \citet{get07} point out, when metals escape into the IGM they preferentially migrate 
into voids of low density and do not influence star formation until they either recollapse 
into the halo or merge with a different one.  However, the remnant itself could also be 
prone to breakup and clumping after colliding with the \ion{H}{2} region shell, possibly 
triggering a second prompt generation of star formation in the enriched gas.  These stars 
would reside in the outer regions of the halo, at radii of 100 - 200 pc.  The implications 
of this novel mechanism for global star formation at high redshifts remain to be 
investigated with three-dimensional models that include metal line cooling.  Type II 
and pair instability supernovae would imprint very different nucleosynthetic signatures 
on their descendants, which is important to surveys of ultra and extremely metal-poor 
stars in the galactic halo.

Immediate low-mass star formation is probable when metals are confined to the halo,
given the short cooling and free-fall times in gas at densities of 10$^4$ - 10$^5$
cm$^{-3}$ and metallicities of 0.01 solar.  Approximately 10$^{5}$ $\Ms$ of
baryons are enriched to these levels, and it is conceiveable that this gas could
cool and fracture into a swarm of less-massive and long-lived stars that are 
confined to radii of 10 - 20 pc by the dark matter potential of the halo.  Star 
formation would be complicated by the presence of a central black hole with large 
x-ray luminosities fueled by fallback into the halo.  Nonetheless, this scenario 
raises the intriguing possibility of globular cluster formation at high redshifts,
various mechanisms for which have been studied in the past 40 years \citep{pd68,p84,kg05}.

\section{Discussion and Conclusions}

Primordial supernovae can be more damaging to cosmological minihalos than previous 
studies suggest when they are properly initialized with kinetic rather than thermal 
energy.  This is primarily because the momentum of the ejecta cannot be radiated away, 
even when radiative cooling is highly efficient, as in partially ionized or neutral 
clouds.  Even blasts that are trapped by massive halos severely disrupt them, mixing 
their interiors with metals and creating massive fallback onto any black hole that 
remains.  Population 
III stars of lower mass that are capable of explosions (15 - 40 $\Ms$) easily destroy 
halos $\lesssim$ 10$^7$ $\Ms$ but not more massive ones, in part because they cannot 
preionize them.  On the other hand, primordial stars from 140 - 260 $\Ms$ will destroy 
halos even more massive than 10$^{7}$ $\Ms$ in pair instability supernova blasts.  To a
good approximation, any supernova will destroy any halo less massive than 10$^{7}$ $\Ms$ 
when photoevaporation of the halo prior to the blast is taken into account. 

Primordial supernovae evolve in one of two ways.  Blasts in \ion{H}{2} regions remain free
expansions out to several pc and are decelerated first by adiabatic losses and then later 
by radiative losses.  In addition to the light curve of the radioactive decay of their ejecta, 
these remnants radiate strongly upon collision with the dense shell
of the \ion{H}{2} region, usually by 10$^4$ - 10$^5$ yr.  While most of their luminosity is in
atomic H and He lines, particularly powerful blasts in \ion{H}{2} regions can channel up to 20\% 
of their energy into the CMB.  Explosions in ionized halos evacuate gas from their virial 
radii to densities below 10$^{-4}$ cm$^{3}$.  Our models are in qualitative agreement with 
earlier one-dimensional Lagrangian and three-dimensional smoothed particle hydrodynamics 
(SPH) calculations of Type II and pair instability SN in primordial \ion{H}{2} regions.  Our 
results are also consistent with early one-dimensional calculations of supernova blasts
in uniform diffuse media \citep{chev74,ciof88}:  they too manifest reverse shocks that 
separate from forward shocks at relatively small radii and exhibit multiple shock 
reflections throughout the interior.

On the other hand, blasts in nearly neutral halos promptly radiate most of their kinetic 
energy as x-rays but retain sufficient momentum for the ejecta to reach radii of 10 - 20
pc.  These contained explosions leave little observational footprint but radically alter 
the evolution of gas within the halo, first by contamination with heavy elements and 
then by successive cycles of fallback and rebound.  This is a departure from earlier
supernovae models in neutral halos initialized with thermal rather than kinetic energy
that is radiated away before launching any flow out into the halo.  Again, in agreement
with earlier work by \citet{chev74} and \citet{ciof88}, we find that neither explosion 
pathway is well represented by the canonical phases of idealized remnants in uniform media 
(Sedov-Taylor, PDS, or MCS).  In all cases inclusion of the dark matter potential is crucial; 
models would otherwise fail to capture fallback in trapped explosions or erroneously predict 
expansion of the remnant until it achieves pressure equilibrium with the IGM.

Primordial supernovae occurred in neutral halos only if the halo was $\gtrsim$ 10$^7 
\Ms$ and the progenitor was not very massive.  Given that stars normally form by H$_2$ 
cooling well before the halo reaches this mass, is it likely that such blasts ever 
happened?  The answer is yes if one considers halos at slightly later redshifts in
rising Lyman-Werner backgrounds, whose effect is to postpone star formation in halos 
until they reach larger masses and become self-shielded to photodissociating photons
\citep{mac01,oshea07a,wa07c}.  Numerical surveys indicate that when stars form in more 
massive halos whose virial temperatures approach the threshold for atomic line cooling, 
they still do so by molecular hydrogen cooling and are subject to the range in masses 
examined in this paper \citep{oshea07b}.  It is probable that explosions and fallback in 
more massive halos were more common at somewhat lower redshifts (15 - 20) than those of 
the very first stars.  We also point out that there are a greater range of spin parameters 
for star-forming halos than are sampled in current studies.  Halos with higher angular 
momenta exhibit markedly lower central infall rates \citep{oet05} that lead to much less 
massive stars.  Finally, since the processes that halt accretion onto primordial stars
remains poorly understood, low-mass supernovae in relatively massive halos cannot be ruled 
out. 
  
The recent discovery of the first potential pair instability supernova ever observed 
\citep{ns07} lends hope that this sort of explosion may have occurred in the high 
redshift universe.  Population III PISN might be seen at moderate redshift ($z \sim$ 
6) with current technology \citep{scan05}, or out to significantly higher redshift 
with the \textit{James Webb Space Telescope} \citep{wl05}.  Scannapieco et al. use 
the KEPLER code to create mock supernova light curves and demonstrate that the 
typical luminosities of pair instability supernovae are not much brighter than 
standard Type Ia supernovae.  However, they remain brighter for much longer periods 
of time and have distinctive spectral features, making them easy to discriminate from  
standard Type Ia or Type II supernovae.  Unfortunately, time dilation may make the 
light curves of extremely high redshift supernovae difficult to detect, especially
those powered by the pair instability, since they have long evolutionary times in
the rest frame.  Their observation would require very different survey strategies
than those employed in low redshift supernova searches.  Furthermore, the predicted 
Population III supernova rate varies widely depending on a range of basic assumptions 
\citep{wl05,wa05,mes06}.

If Population III supernovae are ever observed, they may be useful probes of their 
immediate environments.  Peculiar features in a supernova light curve can be exploited
to infer the properties of circumstellar gas \citep{sm07}.  If Population III stars are 
rapidly rotating, they may explode as gamma ray bursts \citep{bl06} and could be used 
to probe the interstellar medium of the halos in which they form, as has been done in 
the $z <$ 4 universe \citep{cet05,pet06,pet07}.

The ionization front of the progenitor itself would also be prone to violent dynamical
instabilities in three dimensions due to H$_2$ cooling, forming gas clumps in the relic 
\ion{H}{2} region with which the blast would interact \citep{wn07a,wn07b}.  They would be most 
prominent in \ion{H}{2} regions trapped in massive halos, partly because D-type fronts better 
shield H$_2$ from Lyman-Werner photons from the star and partly because R-type I-fronts 
that break free of halos would tend to photoevaporate any clumps that may have formed, 
although numerical simulations indicate that some would survive \citep{wn07a,wn07b}.  
I-front instabilities may also drive turbulence in the surrounding medium.  Both 
fragmentation and turbulence will enhance the mixing of metals in halos and should be 
included for self-consistency in future three dimensional models.

Molecular hydrogen and HD could also cause the remnant to cool and fragment, a process 
not addressed in these models.  Explosions in ionized halos would sweep up H$_2$ and HD 
that are rapidly forming in the relic \ion{H}{2} region.  However, molecular cooling would 
not occur until late times in our models because the remnant remains fully ionized and 
hot for several Myr, as shown in Figure \ref{fig:260Ms_halo1_2}f.  At 7.93 Myr, H$_2$ might 
form in the forward shock as it partially recombines and cools to 10000 K, forming a 
dense shell that might be prone to breakup.  In all likelihood, prior instabilities,
such as Rayleigh-Taylor instabilities in the shock,
would pre-empt those due to H$_2$ cooling, but the radiative losses might 
still be relevant to the kinematics of the remnant.  Molecular hydrogen cooling might
also be viable in trapped explosions.  Figures \ref{fig:40Ms_halo3_2}b and f show that the 
leading shock in the contained blast recombines and cools at much earlier times, at about 
the same time instabilities would erupt in the contact discontinuity.  Rayleigh-Taylor
instabilities in the reverse shock and Vishniac instabilities in the forward shock
mediated by H$_2$ cooling could evolve together, destroying the shell from within and 
without.

In the absence of magnetic fields in the vicinity of the progenitor, how would the free 
expansion interact with the \ion{H}{2} region or neutral halo?  Simple collisional cross 
section arguments imply that the ejecta particles would travel to great distances with 
little deflection by the surrounding gas unless magnetic fields mediate hydromagnetic 
interactions between the blast and its environment.  If no primordial fields were present, 
the ejecta might interact differently with the halo than predicted by the fluid equations, 
which are predicated upon mean free paths in the flow being far shorter than characteristic 
lengths in the flow itself.  Nevertheless, for now it is reasonable to suppose that magnetic 
fields arise due to the Weibel instability \citep[e.~g. ][]{md07}, effectively coupling the 
two plasma streams.  This mechanism has been invoked to explain the origin of GRB afterglows 
\citep{mlm07}.

An important phenomenon not noted in earlier surveys of primordial supernovae are the
large central infall rates that can result from fallback when remnants fail to 
escape massive halos, in excess of 10$^{-2}$ $\Ms$ yr$^{-1}$ and lasting 1 - 2 
Myr in some instances.  Such infall, although less coordinated, would certainly appear
in three dimensional simulations and lead to rapid growth of the central black hole.  
The growth would be episodic since bounceback would periodically reverse the inflow,  
and three dimensional radiation hydrodynamical accretion simulations would be required  
to ascertain how much gas at the center of the halo would fall into the black hole. 
Nevertheless, this mechanism may provide the means whereby black holes of several 
thousand solar masses could form in pregalactic structures at $z \sim$ 20, which
numerical studies suggest could be the origin of the supermassive black holes occupying 
the centers of most massive galaxies today \citep{liet07}.  Periodic accretion onto
the black hole would emit bursts of radiation from the halo that might over time
contribute to an x-ray background at high redshifts and strongly affect the thermal 
and chemical properties of the halo.

One alternative to this scenario is that gas in the trapped explosion instead cools to 
tens of Kelvin by metal-line cooling and then fragments, forming many low mass stars 
whose peculiar motions are confined to the dark matter potential of the halo.  Several 
tens of thousands of solar masses of gas could fuel low mass star formation in this 
manner.  If large numbers of small stars can truly form in the halo the result could be 
a globular cluster 15 - 30 pc in diameter, the radii to which gas in the halo is disturbed 
by the blast.  High dynamical range adaptive mesh refinement (AMR) simulations are necessary 
to determine if gas in the halo can fracture and collapse in this fashion.

Finally, our supernova models may hint at an alternative pathway for the assembly of the
first primitive galaxies.  \citet{get07} find that the first supernovae preferentially 
expel metals into the voids interspersed among cosmological filaments.  Since densities 
are relatively low there, the metals do not promote immediate star formation, and must 
generally await merger timescales to be taken up into a new generation of stars.  If 
supernovae ejecta instead mix efficiently with the dense shells of relic H II regions and 
cause them to cool and fragment, a second generation of stars with different chemical 
abundance patterns might promptly form.  Likewise, explosions trapped by massive halos
could also create small populations of new stars before mergers with other halos.  If so, 
the first primitive galaxies were assembled sooner, with more stars and different chemical 
abundances than in current cosmological models.  Upcoming observations by the \textit{James 
Webb Space Telescope} and \textit{Thirty Meter Telescope} might discriminate between these
two scenarios.
   
\acknowledgments

We thank the anonymous referee, whose detailed comments significantly improved the quality 
of this work.  We also thank Ken Nomoto for providing us his hypernova yields and 
Guillermo Garcia-Segura, Alex Heger, and Lars Bildsten for helpful discussions 
concerning these simulations.  BWO was supported by a LANL Director's Postdoctoral 
Fellowship (DOE LDRD grant 20051325PRD4).  This work was carried out under the 
auspices of the National Nuclear Security Administration of the U.S. Department of 
Energy at Los Alamos National Laboratory under Contract No. DE-AC52-06NA25396.  The 
simulations were performed on the open cluster Coyote at Los Alamos National 
Laboratory.


\begin{thebibliography}{}

\bibitem[Abel et al.(2000)]{abn00} Abel, T., Bryan, G.~L., \& Norman, M.~L.
\ 2000, \apj, 540, 39 
\bibitem[Abel et al.(2002)]{abn02} Abel, T., Bryan, G.~L., \& Norman, M.~L.
\ 2002, Science, 295, 93
\bibitem[Abel et al.(2007)]{awb07} Abel, T., Wise, J.~H., \& Bryan, G.~L.\ 2007, \apjl, 
659, L87 
\bibitem[Abel et al.(1997)]{abet97} Abel, T., Anninos, P., Zhang, Y., \& Norman, M.~L.\ 
1997, New Astronomy, 2, 181 
\bibitem[Alvarez et al.(2006)]{abs06} Alvarez, M.~A., Bromm, 
V., \& Shapiro, P.~R.\ 2006, 
\apj, 639, 621 
\bibitem[Anninos et al.(1997)]{anet97} Anninos, P., Zhang, Y., Abel, T., \& Norman, 
M.~L.\ 1997, New Astronomy, 2, 209 
\bibitem[Arnett(1982)]{a82} Arnett, W.~D.\ 1982, \apj, 253, 785 
\bibitem[Baraffe, Heger, \& Woosley(2001)]{bhw01} Baraffe, 
I., Heger, A., \& Woosley, S.~E.\ 2001, \apj, 550, 890 
\bibitem[Branch \& Tammann(1992)]{bt92} Branch, D., \& Tammann, G.~A.\ 1992, \araa, 30, 359 
\bibitem[Bromm et al.(1999)]{bcl99} Bromm, V., Coppi, P.~S., \& Larson, 
R.~B.\ 1999, \apjl, 527, L5 
\bibitem[Bromm et al.(2001)]{bcl01} Bromm, V., Ferrara, A., Coppi, P.~S., \& Larson, R.~B.\ 
2001, \mnras, 328, 969 
\bibitem[Bromm et al.(2003)]{byh03} Bromm, V., Yoshida, N., \& Hernquist, L.\ 2003, \apjl, 596, 
L135 
\bibitem[Bromm \& Loeb(2006)]{bl06} Bromm, V., \& Loeb, A.\ 2006, \apj, 642, 382 
\bibitem[Chen et al.(2005)]{cet05} Chen, H.-W., Prochaska, J.~X., Bloom, J.~S., \& Thompson, 
I.~B.\ 2005, \apjl, 634, L25 
\bibitem[Chen \& Miralda-Escud{\'e}(2004)]{cm04} Chen, X., \& Miralda-Escud{\'e}, J.\ 2004, 
\apj, 602, 1 
\bibitem[Chevalier(1974)]{chev74} Chevalier, R.~A.\ 1974, \apj, 188, 501 
\bibitem[Chevalier \& Imamura(1982)]{ci82} Chevalier, R.~A., \& Imamura, J.~N.\ 1982, \apj, 
261, 543 
\bibitem[Chugai \& Chevalier(2006)]{cc06} Chugai, N.~N., \& Chevalier, R.~A.\ 2006, \apj, 641, 
1051 
\bibitem[Churchwell(2002)]{church02} Churchwell, E.\ 2002, \araa, 40, 27 
\bibitem[Cioffi et al.(1988)]{ciof88} Cioffi, D.~F., McKee, C.~F., \& Bertschinger, E.\ 1988, 
\apj, 334, 252 
\bibitem[Cox(1972)]{cox72} Cox, D.~P.\ 1972, \apj, 178, 159 
\bibitem[Dwarkadas(2005)]{dw05} Dwarkadas, V.~V.\ 2005, \apj, 630, 892 
\bibitem[Franco et al.(1990)]{ftb90} Franco, J., Tenorio-Tagle, G., \& Bodenheimer, P.\ 1990, 
\apj, 349, 126 
\bibitem[Galli \& Palla(1998)]{gp98} Galli, D., \& Palla, F.\ 1998, \aap, 335, 403 
\bibitem[Gao et al.(2007)]{gao07} Gao, L., Yoshida, N., Abel, T., Frenk, C.~S., Jenkins, A., 
\& Springel, V.\ 2007, \mnras, 378, 449 
\bibitem[Greif et al.(2007)]{get07} Greif, T.~H., Johnson, J.~L., Bromm, V., \& Klessen, R.~S.
\ 2007, \apj, 670, 1 
\bibitem[Hayes et al.(2006)]{het06} Hayes, J.~C., Norman, M.~L., Fiedler, R.~A., Bordner, 
J.~O., Li, P.~S., Clark, S.~E., ud-Doula, A., \& Mac Low, M.-M.\ 2006, \apjs, 165, 188 
\bibitem[Heger \& Woosley(2002)]{hw02} Heger, A.~\& Woosley, S.~E.\ 2002, \apj, 567, 
532 
\bibitem[Imamura et al.(1984)]{iet84} Imamura, J.~N., Wolff, M.~T., \& Durisen, R.~H.\ 1984, 
\apj, 276, 667 
\bibitem[Johnson et al.(2007)]{jet07} Johnson, J.~L., Greif, T.~H., \& Bromm, V.\ 2007, 
\apj, 665, 85 
\bibitem[Kitayama \& Yoshida(2005)]{ky05} Kitayama, T., \& Yoshida, N.\ 2005, \apj, 630, 675 
\bibitem[Kitayama et al.(2004)]{ket04} Kitayama, T., Yoshida, 
N., Susa, H., \& Umemura, M.\ 2004, \apj, 613, 631 
\bibitem[Kravtsov \& Gnedin(2005)]{kg05} Kravtsov, A.~V., \& Gnedin, O.~Y.\ 2005, \apj, 623, 650 
\bibitem[Kudritzki(2000)]{kud00} Kudritzki, R.\ 2000, The 
First Stars.~Proceedings of the MPA/ESO Workshop held at Garching, Germany, 
4-6 August 1999.~Achim Weiss, Tom G.~Abel, Vanessa Hill (eds.).~Springer, 
\bibitem[Li et al.(2007)]{liet07} Li, Y., et al.\ 2007, \apj, 665, 187 
\bibitem[Machacek et al.(2001)]{mac01} Machacek, M.~E., Bryan, G.~L., \& Abel, T.\ 2001, \apj, 
548, 509 
\bibitem[Mackey et al.(2003)]{mbh03} Mackey, J., Bromm, V., \& Hernquist, L.\ 2003, \apj, 586, 1 
\bibitem[Madau et al.(2001)]{met01} Madau, P., Ferrara, A., \& Rees, M.~J.\ 2001, \apj, 555, 92 
\bibitem[McKee \& Ostriker(1977)]{mo77} McKee, C.~F., \& Ostriker, J.~P.\ 1977, \apj, 218, 148 
\bibitem[Medvedev(2007)]{md07} Medvedev, M.~V.\ 2007, \apss, 307, 245 
\bibitem[Medvedev et al.(2007)]{mlm07} Medvedev, M.~V., Lazzati, D., Morsony, B.~C., \& Workman, 
J.~C.\ 2007, \apj, 666, 339 
\bibitem[Mesinger et al.(2006)]{mes06} Mesinger, A., Johnson, B.~D., \& Haiman, Z.\ 2006, \apj, 
637, 80 
\bibitem[Meynet et al.(2007)]{met07} Meynet, G., Ekstrom, S., Maeder, A., Hirschi, R., 
Chiappini, C., \& Georgy, C.\ 2007, ArXiv e-prints, 709, arXiv:0709.2275 
\bibitem[Oh et al.(2003)]{ock03} Oh, S.~P., Cooray, A., \& Kamionkowski, M.\ 2003, \mnras, 342, 
L20 
\bibitem[Oort(1951)]{oort51} Oort, J.~H.\ 1951, in \textit{Problems of Cosmical Aerodynamics}
(Dayton, Ohio: Central Air Document Office), p. 118.
\bibitem[O'Shea et al.(2005)]{oet05} O'Shea, B.~W., Abel, T., Whalen, D., \& Norman, M.~L.\ 2005, 
\apjl, 628, L5 
\bibitem[O'Shea \& Norman(2007)]{oshea07a} O'Shea, B.~W., \& Norman, M.~L.\ 2007, \apj, 673, 14
\bibitem[O'Shea \& Norman(2007)]{oshea07b} O'Shea, B.~W., \& Norman, M.~L.\ 2007, \apj, 654, 66 
\bibitem[O'Shea et al.(2004)]{enzom} O'Shea, B.~W., Bryan, G., Bordner, J., Norman, M.~L., Abel, 
T., Harkness, R., \& Kritsuk, A.\ 2004, ArXiv Astrophysics e-prints, arXiv:astro-ph/0403044 
\bibitem[Peebles(1984)]{p84} Peebles, P.~J.~E.\ 1984, \apj, 277, 470 
\bibitem[Peebles \& Dicke(1968)]{pd68} Peebles, P.~J.~E., \& Dicke, R.~H.\ 1968, \apj, 154, 891 
\bibitem[Pinto \& Eastman(2001)]{pe01} Pinto, P.~A., \& Eastman, R.~G.\ 2001, New Astronomy, 6, 
307 
\bibitem[Prochaska et al.(2006)]{pet06} Prochaska, J.~X., Chen, H.-W., \& Bloom, J.~S.\ 2006, 
\apj, 648, 95 
\bibitem[Prochaska et al.(2007)]{pet07} Prochaska, J.~X., Chen, H.-W., Dessauges-Zavadsky, M., 
\& Bloom, J.~S.\ 2007, \apj, 666, 267 
\bibitem[Qiu et al.(2008)]{qiu08} Qiu, J.-M., Shu, C.-W., Liu, J.-R., \& Fang, L.-Z.\ 2008, 
New Astronomy, 13, 1 
\bibitem[Ricotti et al.(2001)]{rs01} Ricotti, M., Gnedin, N.~Y., \& Shull, J.~M.\ 2001, \apj, 
560, 580 
\bibitem[Rodr{\'{\i}}guez(2005)]{rod05} Rodr{\'{\i}}guez, L.~F.\ 2005, Massive Star Birth: A 
Crossroads of Astrophysics, 227, 120 
\bibitem[Scannapieco et al.(2005)]{scan05} Scannapieco, E., Madau, P., Woosley, S., Heger, A., 
\& Ferrara, A.\ 2005, \apj, 633, 1031 
\bibitem[Schaerer(2002)]{s02} Schaerer, D.\ 2002, \aap, 382, 28
\bibitem[Sedov(1959)]{sed59} Sedov, L.~I.\ 1959, Similarity and Dimensional Methods in Mechanics, 
New York: Academic Press, 1959,  
\bibitem[Smith et al.(2007)]{ns07} Smith, N., et al.\ 2007, \apj, 666, 1116 
\bibitem[Smith \& McCray(2007)]{sm07} Smith, N., \& McCray, R.\ 2007, \apjl, 671, L17 
\bibitem[Smith \& Sigurdsson(2007)]{ss07} Smith, B.~D., \& Sigurdsson, S.\ 2007, \apjl, 661, L5 
\bibitem[Stone \& Norman(1992)]{sn92} Stone, J.~M., \& Norman, M.~L.\ 1992, \apjs, 80, 753 
\bibitem[Tammann \& Schroeder(1990)]{ts90} Tammann, G.~A., \& Schroeder, A.\ 1990, \aap, 236, 149 
\bibitem[Taylor(1950)]{tay50} Taylor, G.~I.\ 1950, \textit{Proc.~Roy.~Soc.~London A }, 201,
159  
\bibitem[Tegmark et al.(1993)]{tse93} Tegmark, M., Silk, J., \& Evrard, A.\ 1993, \apj, 417, 54 
\bibitem[Tominaga et al.(2007)]{tun07} Tominaga, N., Umeda, H., \& Nomoto, K.\ 2007, \apj, 660, 516 
\bibitem[Truelove \& McKee(1999)]{tm99} Truelove, J.~K., \& McKee, C.~F.\ 1999, \apjs, 120, 299 
\bibitem[Vink, de Koter, \& Lamers(2001)]{vkl01} Vink, J.~S., de Koter, A., \& Lamers, H.~J.~G.~L.
~M.\ 2001, \aap, 369, 574 
\bibitem[Voit(1996)]{v96} Voit, G.~M.\ 1996, \apj, 465, 548 
\bibitem[Weinmann \& Lilly(2005)]{wl05} Weinmann, S.~M., \& Lilly, S.~J.\ 2005, \apj, 624, 526 
\bibitem[Whalen et al.(2004)]{wan04} Whalen, D., Abel, T., \& Norman, M.~L.\ 2004, \apj, 610, 14 
\bibitem[Whalen \& Norman(2006)]{wn06} Whalen, D., \& Norman, M.~L.\ 2006, \apjs, 162, 281 
\bibitem[Whalen \& Norman(2007a)]{wn07a} Whalen, D.~J., \& Norman, M.~L.\ 2007, ArXiv Astrophysics 
e-prints, arXiv:astro-ph/0703463 
\bibitem[Whalen \& Norman(2007b)]{wn07b} Whalen, D., \& Norman, M.~L.\ 2007, ArXiv e-prints, 708, 
arXiv:0708.2444 
\bibitem[Wise \& Abel(2005)]{wa05} Wise, J.~H., \& Abel, T.\ 2005, \apj, 629, 615 
\bibitem[Wise \& Abel(2007a)]{wa07a} Wise, J.~H., \& Abel, T.\ 2007, ArXiv e-prints, 710, 
arXiv:0710.3160 
\bibitem[Wise \& Abel(2007b)]{wa07b} Wise, J.~H., \& Abel, T.\ 2007, ArXiv e-prints, 710, 
arXiv:0710.4328 
\bibitem[Wise \& Abel(2007c)]{wa07c} Wise, J.~H., \& Abel, T.\ 2007, \apj, 671, 1559
\bibitem[Yoshida et al.(2007)]{yet07} Yoshida, N., Oh, S.~P., Kitayama, T., \& Hernquist, L.\ 
2007, \apj, 663, 687 


\end{thebibliography}
\end{document}